# Feynman Rules for Simplicial Gravity

*Herbert W. Hamber* * and *Shao Liu* †

Department of Physics
University of California at Irvine
Irvine, CA 92717

## ABSTRACT

We develop the general formalism for performing perturbative diagrammatic expansions in the lattice theory of quantum gravity. The results help establish a precise correspondence between continuum and lattice quantities, and should be a useful guide for non-perturbative studies of gravity. The Feynman rules for Regge's simplicial lattice formulation of gravity are then discussed in detail in two dimensions. As an application, the two-dimensional conformal anomaly due to a $D$-component scalar field is explicitly computed in perturbation theory.

*e-mail address : hhamber@uci.edu
†e-mail address : lius@uci.edu

# 1 Introduction

In the quantization of gravitational interactions one expects non-perturbative effects to play an important role. One formulation available for studying such effects is Regge's simplicial lattice theory of gravity [1]. It is the only lattice model with a local gauge invariance [2], and the only model known to contain gravitons in four dimensions [3]. A number of fundamental issues in quantum gravity, such as the existence of non-trivial ultraviolet fixed points of the renormalization group in four dimensions and the recovery of general relativity at large distances, can in principle be addressed in such a model. The presence of a local gauge invariance, which is analogous to the diffeomorphism group in the continuum, makes the model attractive as a regulated theory of gravity [4], while the existence of a phase transition in three and four dimensions [5,6,7,8] (but not in two [9,10]) suggests the existence of a (somewhat unusual) lattice continuum limit. A detailed discussion of the properties of the two phases characterizing four-dimensional gravity, and of the associated critical exponents, can be found in [7]. Recently calculations have progressed to the point that a first calculation of the Newtonian potential from the correlation of heavy particle world lines, following the proposal of [11], has become feasible [12]. These results indicate that in the lattice quantum theory the potential is indeed attractive, and has the correct heavy mass dependence. In the same work a general scaling theory for gravitational correlations, valid in the vicinity of the fixed point, was put forward.

In view of this recent progress it would appear desirable to further elucidate the correspondence between continuum and lattice theories. The weak field expansion is available to systematically develop this correspondence, and it is well known that such an expansion can be carried out in both formulations. Not unexpectedly, it is technically somewhat more complex in the lattice theory due to the presence of additional vertices, as happens in ordinary lattice gauge theories. In the past most perturbative studies of lattice gravity have focused on the lowest order terms, and in particular the lattice graviton propagators [3,9,13]. As such, these did not probe directly important, genuinely quantum-mechanical, aspects of the theory. A systematic weak field expansion is generally useful, since it allows one in principle to determine subleading lattice corrections to the continuum results, which can be relevant in the analysis of the numerical non-perturbative results in the full theory.

More importantly, the weak field expansion can be used to compare with known results in the continuum, and some are known in two dimensions [14,15]. A related motivation comes from trying to understand the recently discovered discrepancy between the critical exponents for matter coupled to gravity in two dimensions as computed in the lattice regularized model for gravity [16,17], and the corresponding conformal field theory predictions [18,19]. Particularly significant in this respect appears to be the recent realization that the conformal field theory exponents describe two-dimensional random systems in *flat* space, and do not correspond to "gravitational" dressing of correlators [21,22].

The plan of the paper is as follows. We shall first introduce our notation and describe the gravitational action, including matter fields. In order to keep our discussion as simple as possible, we shall limit it mostly to the two-dimensional case, although it can be easily generalized to higher dimensions. The subsequent sections will then be devoted to the systematic development of the lattice weak field expansion. The results presented here will help elucidate the correspondence between the lattice and continuum theories, and bring out the role of local gauge invariance in the lattice theory. We shall then discuss the Feynman rules for gravity coupled to a scalar field, and as an application compute the conformal anomaly in two dimensions. The final section contains our conclusions.

# 2 The Discretized Theory

In this section we shall briefly review the construction of the action describing the gravitational field on the lattice, and use the occasion to define the notation used later in the paper. In concrete examples we will often refer, because of its simplicity, to the two-dimensional case, where a number of results can be derived easily and transparently. In a number of instances though important aspects of the discussion will be quite general, and not restricted to specific aspects of the two-dimensional case.

It is well known that in two dimensions quantum gravity can be defined on a two-dimensional surface consisting of a network of flat triangles. The underlying lattice may be constructed in a number of ways. Points may be distributed randomly on the surface and then joined to form triangles according to some algorithm. An alternative procedure is to start with a regular lattice, like a tessellation of the two sphere or a lattice of squares divided into triangles by drawing in parallel sets of diagonals, and then allow the edge lengths to vary, which will introduce curvature localized on the vertices. For arbitrary assignments of edge lengths, consistent with the imposition of the triangle inequalities constraints, such a lattice is in general far from regular, and resembles more a random lattice. In the following though we will narrow down the discussion, and think of the "regular" lattice as consisting of a network of triangles with a fixed coordination number of six, although many of the results in this work are expected to be quite general and should not depend significantly on the specific choice of local coordination numbers.

The elementary degrees of freedom on the lattice are the edge lengths, with the correspondence between continuum and lattice degrees of freedom given locally by

$$\{g_{\mu\nu}(x)\}_{x \epsilon \mathcal{M}} \rightarrow \left\{ l_i^2 \right\}_{i=1 \dots N_1} \quad , \tag{2.1}$$

where the index $i$ ranges over all $N_1$ edges in the lattice. From the well known relationship between the induced metric in a simplex and its squared edge lengths,

$$g_{ij}(l^2) = \tfrac{1}{2} \left[ l_{0i}^2 + l_{0j}^2 - l_{ij}^2 \right] \quad . \tag{2.2}$$

one then has the essentially unique functional measure contribution

$$\int \prod_x dg_{\mu\nu}(x) \quad \rightarrow \quad \int_0^\infty \prod_i dl_i^2 \quad , \tag{2.3}$$

supplemented by the additional constraint that the triangle inequalities be satisfied for all quantum fluctuations of the edge lengths [2]. Both the continuum metric and the lattice edge lengths imply some redundancy due to local gauge invariance of the action, which therefore requires gauge-fixing when performing perturbation theory due to the presence of the exact gauge zero modes [3] (the triangle inequalities, which violate gauge invariance by imposing a cutoff on the gauge orbits, are in fact not seen to *any* order in the weak field expansion). ‡ A gauge fixing term is not required in non-perturbative studies, since the contributions from the gauge zero modes are expected to cancel exactly between numerator and denominator in the functional integral for observables, as discussed for example in [4].

For a given action, the dynamics of the lattice will give rise to some average lattice spacing $a_0 = [\langle l^2 \rangle]^{\frac{1}{2}}$, which in turn will supply the ultraviolet cutoff needed to define the quantum theory. It should be stressed that in the following we shall restrict our attention to the *lattice* theory, which is defined in terms of its lattice degrees of freedom *only*. Since it is our purpose to describe an ultraviolet regulated theory of quantum gravity, we shall follow the usual procedure adopted when discussing lattice field theories, and describe, in the spirit of Regge's original idea, the model exclusively in terms of its primary, lattice degrees of freedom: the squared edge lengths. As such, the theory does not require any additional *ad-hoc* regulators for defining conical singularities, for example.

## 2.1 Curvature and Discretized Action

In simplicial gravity the curvature is concentrated on the hinges, which are subspaces of dimensions $d-2$, and is entirely determined from the assignment of the edge lengths. In two dimensions the hinges correspond to the vertices, and $\delta_h$, the deficit angle at a hinge, is defined by

$$\delta_h = 2\pi - \sum_{\substack{\text{triangles } t \\ \text{meeting at } h}} \theta_t \quad , \tag{2.4}$$

where $\theta_t$ is the dihedral angle associated with the triangle $t$ at the vertex $h$ (see Figure 1). In $d$ dimensions several $d$-simplices meet on a $(d-2)$-dimensional hinge, with the deficit angle defined by

$$\delta_h(l^2) = 2\pi - \sum_{\substack{\text{d-simplices} \\ \text{meeting on h}}} \theta_d(l^2) \quad . \tag{2.5}$$

Useful formulas for the cosine of the dihedral angles can be found in [5]. In two dimensions the dihedral angle is obtained from

$$\cos\theta_d = \frac{l_{01}^2 + l_{02}^2 - l_{12}^2}{2 l_{01} l_{02}} \quad . \tag{2.6}$$

(for the labelling see Figure 1).

‡A small breaking of gauge invariance causes well known problems in perturbation theory. There are rather convincing arguments that this is not necessarily the case non-perturbatively, if the breaking can be considered small [24,25], as in the present case

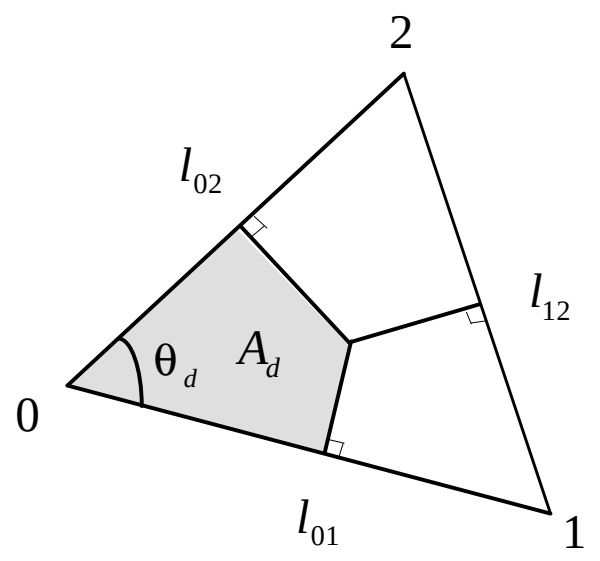

*Fig. 1. Dual area $A_d$ associated with vertex 0, and the corresponding dihedral angle $\theta_d$.*

It is useful to introduce a dual lattice following, for example, the Dirichlet-Voronoi cell construction (see [26,27] and references therein), which consists in introducing perpendicular bisectors in each triangle and joining the resulting vertices. It provides for a natural subdivision of the original lattice in a set of non-overlapping exhaustive cells (see Figure 2.), and furthermore has a natural generalization to higher dimensions.

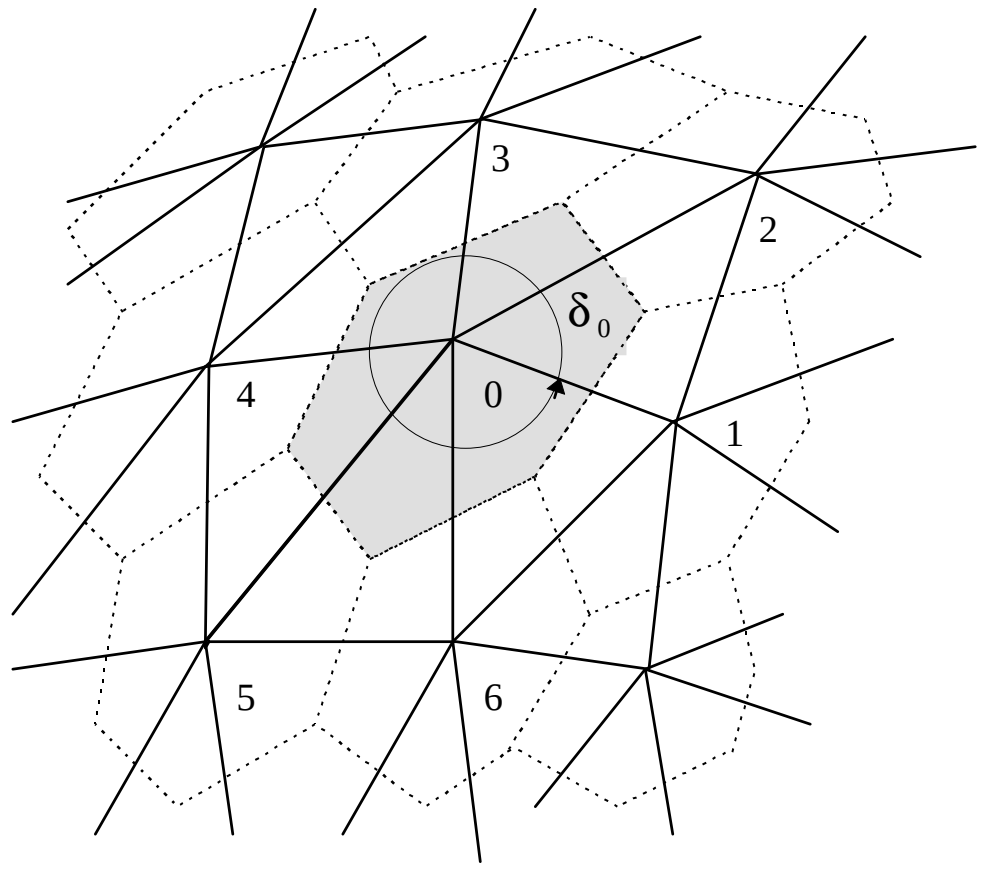

*Fig. 2. Original simplicial lattice (continuous lines) and dual lattice (dotted lines) in two dimensions. The shaded region corresponds to the dual area associated with vertex 0.*

The vertices of the original lattice then reside on circumscribed circles, centered on the vertices of the dual lattice. For the vertex 0 the dihedral dual volume contribution is given by

$$A_d(l^2) = \frac{1}{16A} \left[ l_{12}^2 (l_{01}^2 + l_{02}^2) - (l_{01}^2 - l_{02}^2)^2 \right] \ . \tag{2.7}$$

It should be pointed out that the above subdivision is not unique. Alternatively, one can introduce a baricenter for each triangle, defined as the point equidistant from all three vertices, and again join the resulting vertices. The baricentric dihedral volume is given simply by

$$A_d(l^2) = A/3 \ . \tag{2.8}$$

For the baricentric subdivision one then has simply

$$A_h = \frac{1}{3} \sum_{\substack{\text{triangles t} \\ \text{meeting at h}}} A_t \ . \tag{2.9}$$

$A_h$ can also be taken to be the area of the cell surrounding $h$ in the dual lattice, with

$$A_h = \sum_{\substack{\text{triangles t} \\ \text{meeting at h}}} A_d \ , \tag{2.10}$$

with the dual area contribution for each triangle $A_d$ given in Eq. (2.7). In general, if the original lattice has local coordination number $q_i$ at the site $i$, then the dual cell centered on $i$ will have $q_i$ faces. A fairly complete set of formulae for dual volumes relevant for lattice gravity and their derivation can be found in [5]. In the following we shall refer to the Voronoi cell construction as the "dual subdivision", while we will call the baricentric cell construction the "baricentric subdivision".

It is well known that two-dimensional Einstein gravity is trivial because the Einstein action is constant and the Ricci tensor vanishes identically. When a cosmological constant term and a curvature-squared term are included in the action,

$$I = \int d^2x \sqrt{g} \left[ \lambda - kR + aR^2 \right] \ , \tag{2.11}$$

the classical solutions have constant curvature with $R = \pm\sqrt{\lambda/a}$ (there being no real solutions for $\lambda < 0$). The theory with the Einstein action and a cosmological constant is metrically trivial, having neither dynamical degrees of freedom nor field equations, although non-trivial interactions can arise from the functional measure. The lattice action corresponding to pure gravity is

$$I(l^2) = \sum_h A_h \left[ \lambda - 2k \, R_h + a \, R_h^2 \right] \ , \tag{2.12}$$

with local volume element $A_h$ and the local curvature given by $R_h = 2\delta_h/A_h$. In the limit of small fluctuations around a smooth background, $I(l^2)$ corresponds to the above continuum action [9]. For a manifold of fixed topology the term proportional to $k$ can be dropped, since $\sum_h \delta_h = 2\pi\chi$, where $\chi$ is the Euler characteristic. The curvature-squared leads to non-trivial interactions in two dimensions, although the resulting theory is not unitary.

A number of results have been obtained from the above pure gravitational action. Arguments based on perturbation theory about two dimensions (where the gravitational coupling is dimensionless and the Einstein theory becomes renormalizable) suggest that there should be no non-trivial ultraviolet fixed point of the renormalization group in two dimensions. Explicit calculations in the lattice theory have shown conclusively that this is indeed the case in the absence of matter [9,16,10], as well as in the presence of scalar matter for a sufficiently small number of components [28].

## 2.2 Scalar Field

A scalar field can be introduced, as the simplest type of dynamical matter that can be coupled to the gravitational degrees of freedom. The continuum action is

$$I[g,\phi] = \tfrac{1}{2} \int d^2x \, \sqrt{g} \, [ \, g^{\mu\nu} \, \partial_\mu \phi \, \partial_\nu \phi + (m^2 + \xi R) \phi^2 ] \ , \tag{2.13}$$

The dimensionless coupling $\xi$ is arbitrary. Two special cases are the minimal ($\xi = 0$) and the conformal ($\xi = \frac{1}{6}$) coupling case; in the following we will mostly consider the case $\xi = 0$.

On the lattice consider a scalar $\phi_i$ and define this field at the vertices of the simplices. The corresponding lattice action can be obtained through the usual procedure which replaces the original continuum metric with the induced metric on the lattice, written in terms of the edge lengths [3,13]. Here we shall consider only the two-dimensional case; the generalization to higher dimensions is straightforward. It is convenient to use the notation of Figure 3, which will bring out more readily the symmetries of the resulting lattice action. Here coordinates will be picked in each triangle along the (1,2) and (1,3) directions.

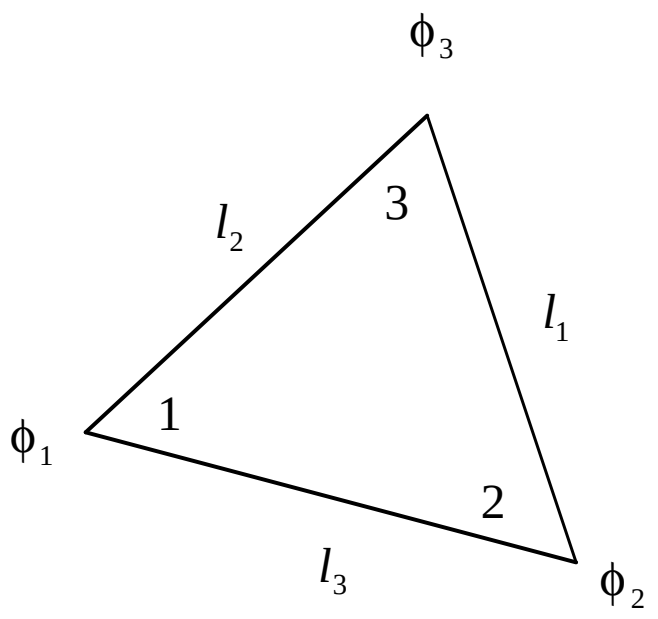

*Fig. 3. Labelling of edges and scalar fields used in the construction of the scalar field action.*

To construct the scalar lattice action, one performs in two dimensions the replacement

$$g_{\mu\nu}(x) \ \longrightarrow \ g_{ij}(\Delta) = \begin{pmatrix} l_3^2 & \frac{1}{2}(-l_1^2 + l_2^2 + l_3^2) \\ \frac{1}{2}(-l_1^2 + l_2^2 + l_3^2) & l_2^2 \end{pmatrix} \ , \tag{2.14}$$

which then gives

$$\det g_{\mu\nu}(x) \ \longrightarrow \ \det g_{ij}(\Delta) = \tfrac{1}{4} \left\{ 2(l_1^2 l_2^2 + l_2^2 l_3^2 + l_3^2 l_1^2) - l_1^4 - l_2^4 - l_3^4 \right\} \equiv 4 A_\Delta^2 \ , \tag{2.15}$$

and also

$$g^{\mu\nu}(x) \ \longrightarrow \ g^{ij}(\Delta) = \frac{1}{\det g(\Delta)} \begin{pmatrix} l_2^2 & \frac{1}{2}(l_1^2 - l_2^2 - l_3^2) \\ \frac{1}{2}(l_1^2 - l_2^2 - l_3^2) & l_3^2 \end{pmatrix} \ . \tag{2.16}$$

For the scalar field derivatives one writes [29,30]

$$\partial_\mu \phi \, \partial_\nu \phi \;\longrightarrow\; \Delta_i \phi \Delta_j \phi = \begin{pmatrix} (\phi_2 - \phi_1)^2 & (\phi_2 - \phi_1)(\phi_3 - \phi_1) \\ (\phi_2 - \phi_1)(\phi_3 - \phi_1) & (\phi_3 - \phi_1)^2 \end{pmatrix} \quad , \tag{2.17}$$

which corresponds to introducing finite lattice differences defined in the usual way by

$$\partial_\mu \phi \;\longrightarrow\; (\Delta_\mu \phi)_i = \phi_{i+\mu} - \phi_i \quad . \tag{2.18}$$

Here the index $\mu$ labels the possible directions in which one can move from a point in a given triangle. The discrete scalar field action then takes the form

$$I(l^2, \phi) \;=\; \tfrac{1}{16} \sum_\Delta \frac{1}{A_\Delta} \left[ l_1^2 (\phi_2 - \phi_1)(\phi_3 - \phi_1) + l_2^2 (\phi_3 - \phi_2)(\phi_1 - \phi_2) + l_3^2 (\phi_1 - \phi_3)(\phi_2 - \phi_3) \right] \quad . \tag{2.19}$$

Using the identity

$$(\phi_i - \phi_j)(\phi_i - \phi_k) = \tfrac{1}{2} \left[ (\phi_i - \phi_j)^2 + (\phi_i - \phi_k)^2 - (\phi_j - \phi_k)^2 \right] \quad , \tag{2.20}$$

one obtains after some re-arrangements the simpler expression [29]

$$I(l^2, \phi) = \tfrac{1}{2} \sum_{<ij>} A_{ij} \left( \frac{\phi_i - \phi_j}{l_{ij}} \right)^2 \quad , \tag{2.21}$$

where $A_{ij}$ is the dual (Voronoi) area associated with the edge $ij$. In terms of the edge length $l_{ij}$ and the dual edge length $h_{ij}$, connecting neighboring vertices in the dual lattice, one has $A_{ij} = \frac{1}{2} h_{ij} l_{ij}$ (see Figure 4). Other choices for the lattice subdivision will lead to a similar formula for the lattice action, but with the Voronoi dual volumes replaced by their appropriate counterparts in the new lattice.

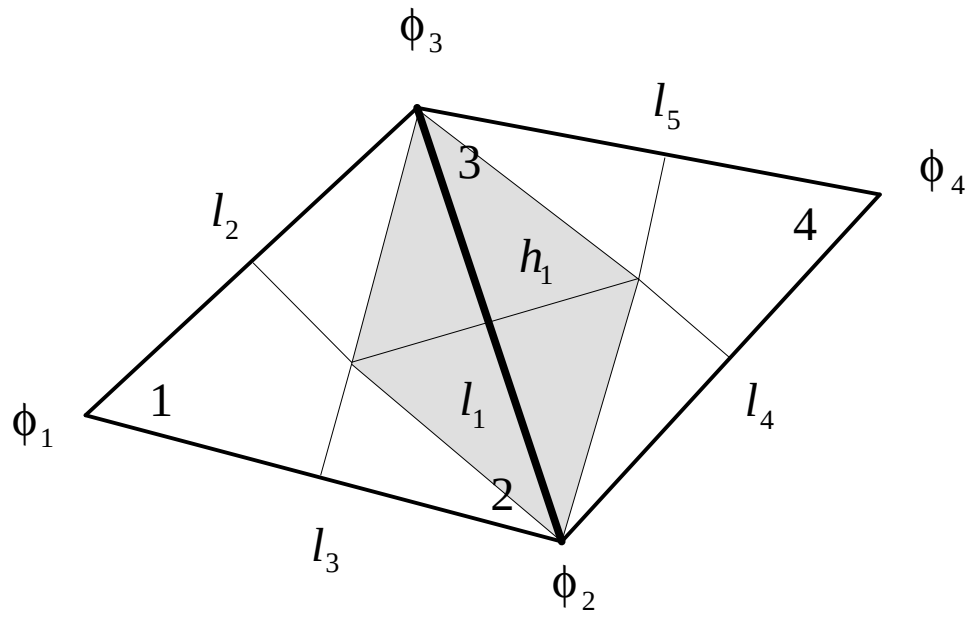

*Fig. 4. Dual area associated with the edge $l_1$ (shaded area), and the corresponding dual link $h_1$.*

For the edge of length 1 the dihedral dual volume contribution is given by

$$A_{l_1} = \frac{l_1^2 (l_2^2 + l_3^2 - l_1^2)}{16 A_{123}} + \frac{l_1^2 (l_4^2 + l_5^2 - l_1^2)}{16 A_{134}} = \tfrac{1}{2} \, l_1 h_1 \quad , \tag{2.22}$$

with $h_1$ is the length of the edge dual to $l_1$. The baricentric dihedral volume for the same edge would be simply

$$A_{l_1} = (A_{123} + A_{134})/3 \quad . \tag{2.23}$$

It is well known that one of the disadvantages of the Voronoi construction is the lack of positivity of the dual volumes, as already pointed out in [5]. Thus some of the weights appearing in Eq. (2.21) can be negative for such an action. On the other hand, for the baricentric subdivision this problem does not arise, as the areas $A_{ij}$ are always positive due to the enforcement of the triangle inequalities. It is immediate to generalize the action of Eq. (2.21) to higher dimensions, with the two-dimensional Voronoi volumes replaced by their higher dimensional analogs.

Mass and curvature terms can be added to the action, so that the total scalar action contribution becomes

$$I(l^2, \phi) = \tfrac{1}{2} \sum_{<ij>} A_{ij} \left( \frac{\phi_i - \phi_j}{l_{ij}} \right)^2 + \tfrac{1}{2} \sum_i A_i \left( m^2 + \xi R_i \right) \phi_i^2 \quad . \tag{2.24}$$

The term containing the discrete analog of the scalar curvature involves the quantity

$$A_i R_i \equiv \sum_{h \supset i} \delta_h \sim \sqrt{g}\, R \quad . \tag{2.25}$$

In the above expression for the scalar action, $A_{ij}$ is the area associated with the edge $l_{ij}$, while $A_i$ is associated with the site $i$. Again there is more than one way to define the volume element $A_i$, [5], but under reasonable assumptions, such as positivity, one expects to get equivalent results in the lattice continuum limit. In the following we shall only consider the simplest form for the scalar action, with $m^2 = \xi = 0$.

# 3 Lattice Weak field Expansion

One of the simplest problems which can be studied analytically in the continuum as well as on the lattice is the analysis of small fluctuations about some classical background solution. In the continuum, the weak field expansion is often performed by expanding the metric and the action about flat Euclidean space

$$g_{\mu\nu} = \delta_{\mu\nu} + \kappa h_{\mu\nu} \quad . \tag{3.1}$$

In four dimensions $\kappa = \sqrt{32\pi G}$, which shows that the weak field expansion there corresponds to an expansion in powers of $G$. In two dimensions this is no longer the case and the relation between $\kappa$ and $G$ is lost; instead one should regard $\kappa$ as a dimensionless expansion parameter which is eventually set to one, $\kappa = 1$, at the end of the calculation. The procedure will be sensible as long as wildly fluctuating geometries are not important in two dimensions (on the lattice or in the continuum). The influence of the latter configurations can only be studied by numerical simulations of the full path integral [9,16].

In the lattice case the weak field calculations can be carried out in three [13] and four [3] dimensional flat background space with the Regge-Einstein action. One finds that the Regge gravity propagator indeed agrees exactly with the continuum result [31] in the weak-field limit.

As a result, the existence of gravitational waves and gravitons in the discrete lattice theory has been established (indeed it is the only lattice theory of gravity for which such a result has been obtained).

In the following we shall consider in detail only the two-dimensional case, although similar calculations can in principle be performed in higher dimensions, with considerable more algebraic effort. In pure gravity case the Einstein-Regge action is a topological invariant in two dimensions, and one has to consider the next non-trivial invariant contribution to the action. We shall therefore consider a two-dimensional lattice with the higher derivative action of Eq. (2.12) and $\lambda = 0$,

$$I(l^2) = 4a \sum_{\text{hinges}\,h} \frac{\delta_h^2}{A_h} \quad . \tag{3.2}$$

The weak field expansion for such a term has largely been done in [9], and we will first recall here the main results. Since flat space is a classical solution for such an $R^2-$type action, one can take as a background space a network of unit squares divided into triangles by drawing in parallel sets of diagonals (see Figure 5). This is one of an infinite number of possible choices for the background lattice, and a rather convenient one. Physical results should in the end be insensitive to the choice of the background lattice used as a starting point for the weak field expansion.

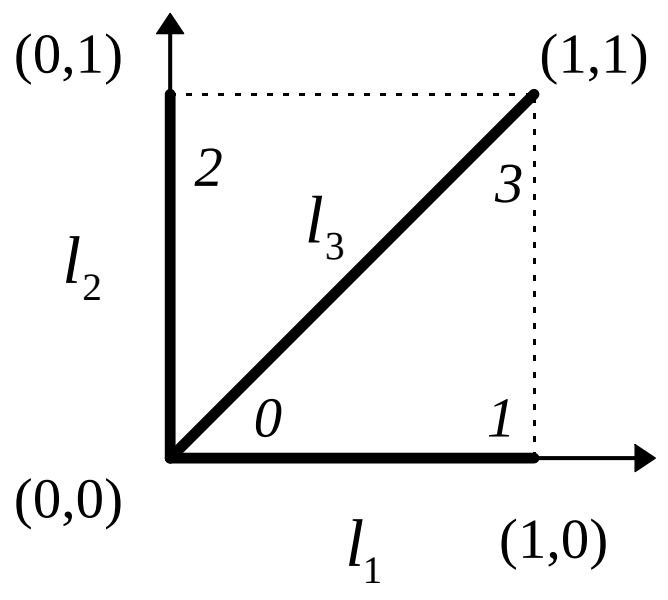

*Fig. 5. Notation for the weak-field expansion about the rigid square lattice.*

It is also convenient to use the binary notation for vertices described in references [3]. As discussed in the previous section, the edge lengths on the lattice correspond to the metric degrees of freedom in the continuum. The edge lengths are thus allowed to fluctuate around their flat space values,

$$l_i = l_i^0 (1 + \epsilon_i) \quad , \tag{3.3}$$

with $l_1^0 = l_2^0 = 1$ and $l_3^0 = \sqrt{2}$ for our choice of lattice. The second variation of the action is then expressed as a quadratic form in the $\epsilon$'s,

$$\delta^2 I = 4a \sum_{ij} \epsilon_i M_{ij} \epsilon_j \quad . \tag{3.4}$$

The properties of $M_{ij}$ are best studied by going to momentum space. One assumes that the fluctuation $\epsilon_i$ at the point $i$, $j$ steps in one coordinate direction and $k$ steps in the other coordinate direction from the origin, is related to the corresponding $\epsilon_i$ at the origin by

$$\epsilon_i^{(j+k)} \; = \; \omega_1^j \, \omega_2^k \, \epsilon_i^{(0)} \quad , \tag{3.5}$$

where $\omega_i = e^{-ik_i}$ and $k_i$ is the momentum in the direction $i$. The matrix $M$ then reduces to a $3 \times 3$ matrix $M_\omega$ with components given by [9]

$$\begin{aligned}
(M_\omega)_{11} &= 2 + \omega_1 - 2\omega_2 - 2\omega_1\omega_2 + \omega_1\omega_2^2 + c.c. \\
(M_\omega)_{12} &= 2 - \omega_1 - \bar{\omega}_2 - \omega_1\omega_2 - \bar{\omega}_1\bar{\omega}_2 - \omega_1^2 - \bar{\omega}_2^2 + \omega_1^2\omega_2 + \bar{\omega}_1\bar{\omega}_2^2 + 2\omega_1\bar{\omega}_2 \\
(M_\omega)_{13} &= 2(-1 + 2\omega_1 - \bar{\omega}_1 + \omega_2 - \bar{\omega}_2 - \omega_1\omega_2 + 2\bar{\omega}_1\bar{\omega}_2 + \bar{\omega}_2^2 - \bar{\omega}_1\bar{\omega}_2^2 - \omega_1\bar{\omega}_2) \\
(M_\omega)_{33} &= 4(2 - 2\omega_1 - 2\omega_2 + \omega_1\omega_2 + \bar{\omega}_1\omega_2 + c.c.)
\end{aligned} \tag{3.6}$$

with the other components easily obtained by symmetry. The change of variables

$$\epsilon_1' = \epsilon_1 \quad \epsilon_2' = \epsilon_2 \quad \epsilon_3' = \tfrac{1}{2}(\epsilon_1 + \epsilon_2) + \epsilon_3 \quad . \tag{3.7}$$

leads for small momenta to the matrix $M_\omega'$ given by

$$M_\omega' \; = \; l^4 \begin{pmatrix} k_2^4 & k_1^2 k_2^2 & -2k_1 k_2^3 \\ k_1^2 k_2^2 & k_1^4 & -2k_1^3 k_2 \\ -2k_1 k_2^3 & -2k_1^3 k_2 & 4k_1^2 k_2^2 \end{pmatrix} + O(k^5) \quad . \tag{3.8}$$

This expression is identical to what one obtains from the corresponding weak-field limit in the continuum theory. To see this, one defines as usual the small fluctuation field $h_{\mu\nu}$ about flat space, which then gives

$$\sqrt{g}\, R^2 = (h_{11,22} + h_{22,11} - 2h_{12,12})^2 + O(h^3) \quad . \tag{3.9}$$

In momentum space, each derivative $\partial_\nu$ produces a factor of $k_\nu$, and one has

$$\sqrt{g}\, R^2 = h_{\mu\nu} \, V_{\mu\nu,\rho\sigma} \, h_{\rho\sigma} \quad , \tag{3.10}$$

where $V_{\mu\nu,\rho\sigma}$ coincides with $M'$ above (when the metric components are re-labelled according to $11 \to 1$, $22 \to 2$, $12 \to 3$).

It is easy to see the reason for the change of variables in Eq. (3.7). Given the three edges in Figure 5, one can write for the metric at the origin

$$g_{ij} = \begin{pmatrix} l_1^2 & \frac{1}{2}(l_3^2 - l_1^2 - l_2^2) \\ \frac{1}{2}(l_3^2 - l_1^2 - l_2^2) & l_2^2 \end{pmatrix} \quad . \tag{3.11}$$

Inserting $l_i = l_i^0(1 + \epsilon_i)$, with $l_i^0 = 1$ for the body principals ($i = 1, 2$) and $l_i^0 = \sqrt{2}$ for the diagonal ($i = 3$), one then obtains

$$\begin{aligned}
l_1^2 = (1 + \epsilon_1)^2 &= 1 + h_{11} \\
l_2^2 = (1 + \epsilon_2)^2 &= 1 + h_{22} \\
\tfrac{1}{2} l_3^2 = (1 + \epsilon_3)^2 &= 1 + \tfrac{1}{2}(h_{11} + h_{22}) + h_{12} \quad ,
\end{aligned} \tag{3.12}$$

which can be inverted to give

$$
\begin{aligned}
\epsilon_1 &= \tfrac{1}{2} h_{11} - \tfrac{1}{8} h_{11}^2 + O(h_{11}^3) \\
\epsilon_2 &= \tfrac{1}{2} h_{22} - \tfrac{1}{8} h_{22}^2 + O(h_{22}^3) \\
\epsilon_3 &= \tfrac{1}{4} (h_{11} + h_{22} + 2h_{12}) - \tfrac{1}{32} (h_{11} + h_{22} + 2h_{12})^2 + O(h^3)
\end{aligned}
\tag{3.13}
$$

which was used in Eq. (3.7). Thus the matrix $M_\omega$ was brought in the continuum form after performing a suitable local rotation from the local edge lengths to the local metric components.

A similar weak field expansion can be performed for the cosmological constant term, although strictly speaking flat space is no longer a classical solution in the presence of such a term [9]. One then obtains a contribution to the second variation of the action of the form

$$
L_\omega = \frac{\lambda}{2} \begin{pmatrix} -1 & 0 & 1 + \bar{\omega}_2 \\ 0 & -1 & 1 + \bar{\omega}_1 \\ 1 + \omega_2 & 1 + \omega_1 & -4 \end{pmatrix} \ . \tag{3.14}
$$

In the weak field limit, and with the same change of variables as described for the matrix $M_\omega$, this leads to

$$
L'_\omega = -\frac{\lambda}{2} \begin{pmatrix} 1 & -1 & 0 \\ -1 & 1 & 0 \\ 0 & 0 & 4 \end{pmatrix} + O(k) \ . \tag{3.15}
$$

The local gauge invariance of the $R^2$-action is reflected in the presence of two exact zero modes in $M_\omega$ of Eq. (3.6). As discussed in Ref. [32], the eigenvalues of the matrix $M_\omega$ are given by

$$
\begin{aligned}
\lambda_1 &= 0 \\
\lambda_2 &= 0 \\
\lambda_3 &= 24 - 9(\omega_1 + \bar{\omega}_1 + \omega_2 + \bar{\omega}_2) + 4(\omega_1 \bar{\omega}_2 + \bar{\omega}_1 \omega_2) \\
&\quad + \omega_1 \omega_2^2 + \omega_1^2 \omega_2 + \bar{\omega}_1 \bar{\omega}_2^2 + \bar{\omega}_1^2 \bar{\omega}_2 \ .
\end{aligned}
\tag{3.16}
$$

It should be noted that the exact zero modes appear for arbitrary $\omega_i$, and not just for small momenta.

It is clear that if one were interested in doing lattice perturbation theory with such an $R^2$ action, one would have to add a lattice gauge fixing term to remove the zero modes, such as the lattice analog of

$$
\frac{1}{\kappa^2} \left( \partial_\mu \sqrt{g(x)} g^{\mu\nu} \right)^2 \ , \tag{3.17}
$$

and then add the necessary Fadeev-Popov nonlocal ghost determinant. A similar term would have to be included as well if one were to pick the lattice analog of the conformal gauge [14]. If one is *not* doing perturbation theory, then of course the contribution of the zero modes will cancel out between the numerator and denominator in the Feynman path integral representation for operator averages, and such a term is not needed, as in ordinary lattice formulations of gauge theories.

The above zero modes correspond to the lattice analogs of diffeomorphisms. It is easy to see that the eigenvectors corresponding to the two zero modes can be written as

$$\begin{pmatrix} \epsilon_1(\omega) \\ \epsilon_2(\omega) \\ \epsilon_3(\omega) \end{pmatrix} = \begin{pmatrix} 1-\omega_1 & 0 \\ 0 & 1-\omega_2 \\ \frac{1}{2}(1-\omega_1\omega_2) & \frac{1}{2}(1-\omega_1\omega_2) \end{pmatrix} \begin{pmatrix} \chi_1(\omega) \\ \chi_2(\omega) \end{pmatrix} , \tag{3.18}$$

where $\chi_1(\omega)$ and $\chi_2(\omega)$ are arbitrary functions (the above result is not restricted to two dimensions; completely analogous zero modes are found for the Regge action in three [13] and four [3] dimensions, leading to expressions rather similar to Eq. (3.18), with $d$ gauge zero modes in $d$ dimensions; their explicit form can be found in the quoted references). In position space one then has

$$\begin{aligned} \epsilon_1(n) &= \chi_1(n) - \chi_1(n+\hat{\mu}_1) \\ \epsilon_2(n) &= \chi_2(n) - \chi_2(n+\hat{\mu}_2) \\ \epsilon_3(n) &= \tfrac{1}{2}\chi_1(n) + \tfrac{1}{2}\chi_2(n) - \tfrac{1}{2}\chi_1(n+\hat{\mu}_1+\hat{\mu}_2) - \tfrac{1}{2}\chi_2(n+\hat{\mu}_1+\hat{\mu}_2) \end{aligned} \tag{3.19}$$

Using the correct relation between induced metric perturbations and edge length variations,

$$\delta g_{ij} = \begin{pmatrix} \delta l_1^2 & \frac{1}{2}(\delta l_3^2 - \delta l_1^2 - \delta l_2^2) \\ \frac{1}{2}(\delta l_3^2 - \delta l_1^2 - \delta l_2^2) & \delta l_2^2 \end{pmatrix} , \tag{3.20}$$

one can easily show that the above corresponds to the discrete analog of the familiar expression

$$\delta g_{\mu\nu} = -\,\partial_\mu \chi_\nu - \partial_\nu \chi_\mu \ , \tag{3.21}$$

which indeed describes the correct gauge variation in the weak field limit. In the discrete case it reflects the invariance of the lattice action under local deformations of the simplicial manifold which leave the local curvatures unchanged [2]. The above relationships express in the continuum the well-known fact that metrics related by a coordinate transformation describe the same physical manifold. Since the continuum metric degrees of freedom correspond on the lattice to the values of edge lengths squared, one expects to find analogous deformations of the edge lengths that leave the lattice geometry invariant, the latter being specified by the local *lattice* areas and curvatures, in accordance with the principle of discussing the geometric properties of the lattice theory in terms of lattice quantities only. This invariance is spoiled by the presence of the triangle inequalities, which places a constraint on how far the individual edge lengths can be deformed. In the perturbative, weak field expansion about a fixed background the triangle inequalities are not seen to any order in perturbation theory, they represent non-perturbative constraints.

The above invariance of the lattice action is a less trivial realization of the exact local gauge invariance found already in one dimension

$$\delta l_n = \chi_{n+1} - \chi_n \ , \tag{3.22}$$

where the $\chi_n$'s represent continuous gauge transformations defined on the lattice vertices [33]. The invariance of the quantum theory is again broken by the triangle inequalities, which in one dimension reduces to the requirement that the edge lengths be positive. Such a breaking is unavoidable in any lattice regularization as it cannot preserve the invariance under scale transformations, which are just special cases of diffeomorphisms.

The weak field expansion for the purely gravitational part can be carried out to higher order, and the Feynman rules for the vertices of order $h^3$, $h^4$, ... in the $R^2$-action of Eq. (2.12) can be derived. Since their expressions are rather complicated, they will not be recorded here.

## 3.1 Feynman Rules

Let us consider next the scalar action of Eq. (2.21),

$$\frac{1}{2}\int d^2x\,\sqrt{g}\,g^{\mu\nu}\,\partial_\mu\phi\,\partial_\nu\phi \;\sim\; \frac{1}{2}\sum_{<ij>} A_{ij}\left(\frac{\phi_i-\phi_j}{l_{ij}}\right)^2 \;. \tag{3.23}$$

In the continuum, the Feynman rules are obtained by expanding out the action in the weak fields $h_{\mu\nu}(x)$,

$$\frac{1}{2}\int d^2x\,\sqrt{g}g^{\mu\nu}\,\partial_\mu\phi\,\partial_\nu\phi \;=\; \frac{1}{2}\int d^2x\,(\partial\phi)^2+\frac{1}{2}\int d^2x\,h_{\mu\nu}\left\{\tfrac{1}{2}\delta_{\mu\nu}(\partial\phi)^2-\partial_\mu\phi\,\partial_\nu\phi\right\}+O(h^2)\;, \tag{3.24}$$

and by then transforming the resulting expressions to momentum space.

On the lattice the action is again expanded in the small fluctuation fields $\epsilon_i$, which depend on the specific choice of parameterization for the flat background lattice. We will first discuss the lattice of Figure 6a, with the choice of labelling according to Figure 5. Figures 6a and 6b correspond to two different gauge choices for the background metric which are physically equivalent; there are many others. The fluctuations in the edge lengths $\epsilon_i$ (see Eq. (3.3)) and the scalar fields $\phi$ at the point $i$, $j$ steps in one coordinate direction and $k$ steps in the other coordinate direction from the origin, are related to the corresponding $\epsilon_i$ and $\phi$ at the origin by

$$\epsilon_i^{(j+k)} \;=\; \omega_1^j\,\omega_2^k\,\epsilon_i^{(0)}\;, \tag{3.25}$$

where $\omega_i = e^{-ik_i}$ and $k_i$ is the momentum in the direction $i$.

$$\phi^{(j+k)} \;=\; {\omega'}_1^j\,{\omega'}_2^k\,\phi_i^{(0)})\;, \tag{3.26}$$

where ${\omega'}_i = e^{-ip_i}$. In practice it is actually more convenient to redefine the edge variables at the midpoints of the links, since this choice neatly removes later a set of complex phase factors from the Feynman rules. For the edge lengths we therefore define the lattice Fourier transforms as

$$\begin{aligned}
\epsilon_1(n) &= \int_{-\pi}^{\pi}\int_{-\pi}^{\pi}\frac{d^2k}{(2\pi)^2}\,e^{-ik\cdot n-ik_1/2}\,\epsilon_1(k)\\
\epsilon_2(n) &= \int_{-\pi}^{\pi}\int_{-\pi}^{\pi}\frac{d^2k}{(2\pi)^2}\,e^{-ik\cdot n-ik_2/2}\,\epsilon_2(k)
\end{aligned}$$

$$\epsilon_3(n) \;=\; \int_{-\pi}^{\pi}\int_{-\pi}^{\pi} \frac{d^2k}{(2\pi)^2}\, e^{-ik\cdot n - ik_1/2 - ik_2/2}\, \epsilon_3(k) \;\;, \tag{3.27}$$

while we still define the scalar field on the vertices, and therefore the Fourier transform in the usual way,

$$\phi(n) = \int_{-\pi}^{\pi}\int_{-\pi}^{\pi} \frac{d^2p}{(2\pi)^2}\, e^{-ip\cdot n}\, \phi(p) \;\;. \tag{3.28}$$

These formulae are easy to generalize to higher dimensions when a simplicial subdivision of a hypercubic lattice is employed, as first suggested in [3].

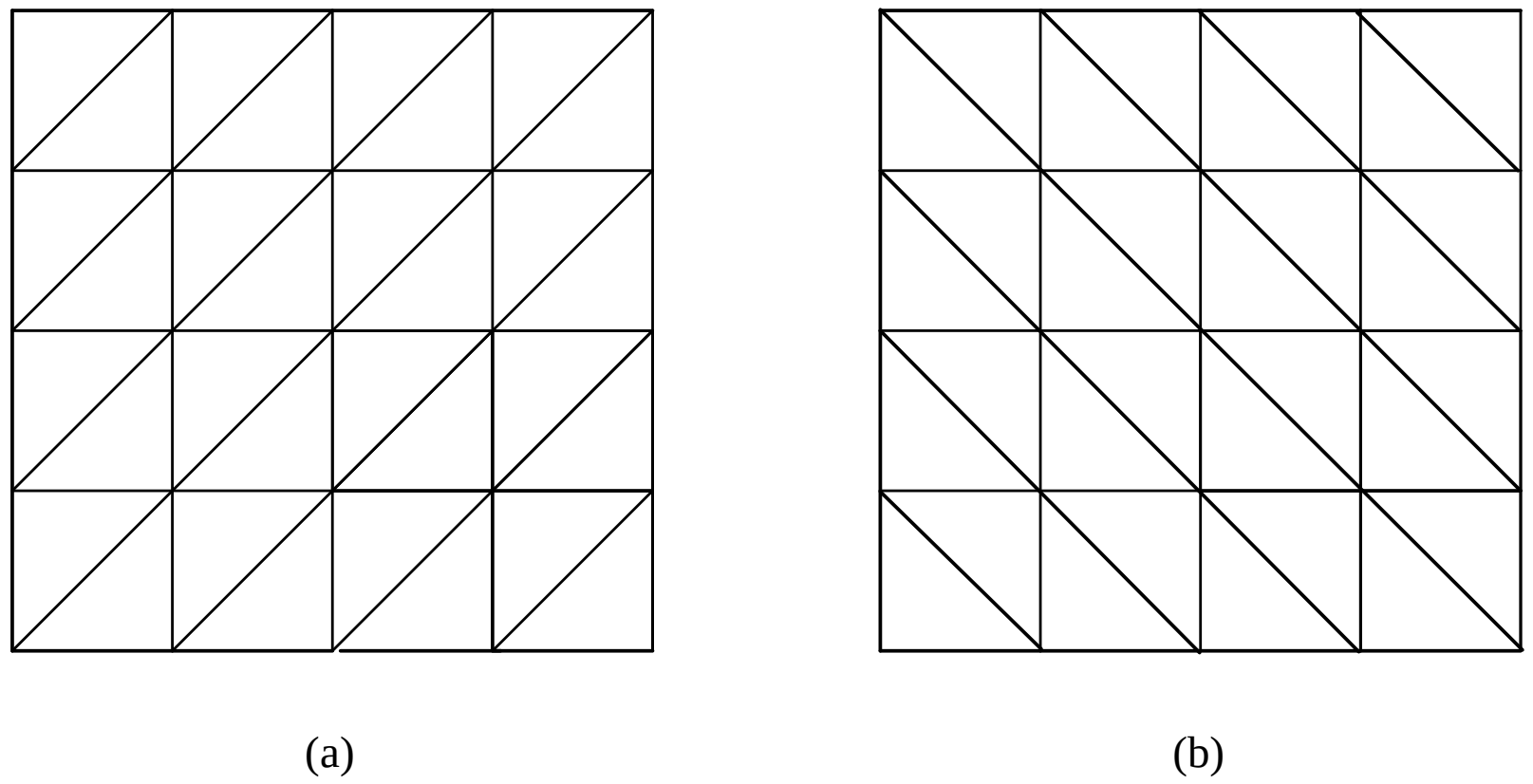

*Fig. 6. Two equivalent triangulation of flat space, based on different subdivisions of the square lattice.*

The kinetic energy term for the scalar field can naturally be decomposed as a sum of three terms

$$\begin{aligned} I \;&=\; \frac{1}{2}\sum_{<ij>} A_{ij}(\frac{\phi_i - \phi_j}{l_{ij}})^2 \\ &=\; \frac{1}{2}\sum_{horizontal} A_{ij}^{(h)}(\frac{\phi_i - \phi_j}{1+\epsilon_1})^2 + \frac{1}{2}\sum_{vertical} A_{ij}^{(v)}(\frac{\phi_i - \phi_j}{1+\epsilon_2})^2 \\ &\quad + \frac{1}{2}\sum_{diagonal} A_{ij}^{(d)}(\frac{\phi_i - \phi_j}{1+\epsilon_3})^2 \;\;, \end{aligned} \tag{3.29}$$

where $< ij >$ labels an edge connecting sites $i$ and $j$. We have separated the sum over all edges into sums over horizontal, vertical, and diagonal edges. The series expansion of each term in the sums with respect to an edge length fluctuation in a particular direction is given by the

following expressions

$$
\begin{aligned}
\frac{A_{ij}^{(h)}}{(1+\epsilon_1)^2} &= \frac{1}{4} - \frac{1}{2}\epsilon_1 + \frac{1}{2}\epsilon_3 \\
&+ \frac{1}{2}\epsilon_1^2 + \frac{1}{4}\epsilon_2^2 + \frac{3}{4}\epsilon_3^2 \\
&+ \frac{1}{2}\epsilon_1\epsilon_2 - \epsilon_1\epsilon_3 - \epsilon_2\epsilon_3 + O(\epsilon_i^3) \ ; \qquad (3.30)
\end{aligned}
$$

$$
\begin{aligned}
\frac{A_{ij}^{(v)}}{(1+\epsilon_2)^2} &= \frac{1}{4} - \frac{1}{2}\epsilon_2 + \frac{1}{2}\epsilon_3 \\
&+ \frac{1}{4}\epsilon_1^2 + \frac{1}{2}\epsilon_2^2 + \frac{3}{4}\epsilon_3^2 \\
&+ \frac{1}{2}\epsilon_1\epsilon_2 - \epsilon_1\epsilon_3 - \epsilon_2\epsilon_3 + O(\epsilon_i^3) \ ; \qquad (3.31)
\end{aligned}
$$

$$
\begin{aligned}
\frac{A_{ij}^{(d)}}{(1+\epsilon_3)^2} &= \frac{1}{4}\epsilon_1 + \frac{1}{4}\epsilon_2 - \frac{1}{2}\epsilon_3 \\
&- \frac{1}{8}\epsilon_1^2 - \frac{1}{8}\epsilon_2^2 - \frac{1}{4}\epsilon_3^2 \\
&- \frac{1}{2}\epsilon_1\epsilon_2 + \frac{1}{2}\epsilon_1\epsilon_3 + \frac{1}{2}\epsilon_2\epsilon_3 + O(\epsilon_i^3) \ . \qquad (3.32)
\end{aligned}
$$

The 0-th order term in $\epsilon$ gives the scalar field propagator

$$
\frac{1}{4\sum_\mu \sin^2(\frac{p_\mu}{2})}, \qquad (3.33)
$$

which is the usual scalar propagator for the square lattice, and coincides with the continuum one for small momenta.

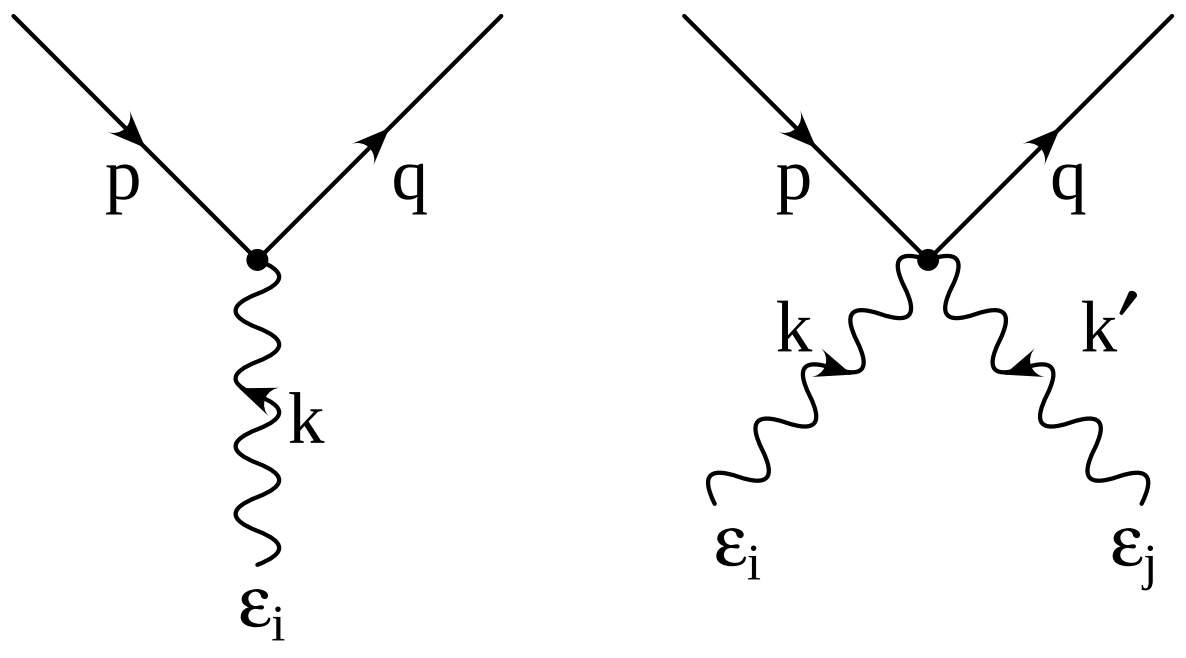

*Fig. 7. Labelling of momenta for the scalar-graviton vertices.*

The higher order terms give, after transforming to momentum space, the Feynman rules for the vertices. For the trilinear vertex associated with $\epsilon_1(k)\phi(p)\phi(q)$ one finds

$$- 2 \sin(\frac{p_1}{2}) \sin(\frac{q_1}{2}) + \cos(\frac{k_2}{2}) \sin(\frac{p_1+p_2}{2}) \sin(\frac{q_1+q_2}{2}) \;; \tag{3.34}$$

for the vertex $\epsilon_2(k)\phi(p)\phi(q)$ one obtains

$$- 2 \sin(\frac{p_2}{2}) \sin(\frac{q_2}{2}) + \cos(\frac{k_1}{2}) \sin(\frac{p_1+p_2}{2}) \sin(\frac{q_1+q_2}{2}) \;; \tag{3.35}$$

and for the vertex $\epsilon_3(k)\phi(p)\phi(q)$ one has

$$\begin{aligned} &2 \cos(\tfrac{k_2}{2}) \sin(\tfrac{p_1}{2}) \sin(\tfrac{q_1}{2}) + 2 \cos(\tfrac{k_1}{2}) \sin(\tfrac{p_2}{2}) \sin(\tfrac{q_2}{2}) \\ &\qquad - 2 \sin(\tfrac{p_1+p_2}{2}) \sin(\tfrac{q_1+q_2}{2}) \;. \end{aligned} \tag{3.36}$$

In order to compare the lattice Feynman rules with the usual continuum ones, one needs to perform a transformation from the $\epsilon_i$ variables to the metric fluctuation $h_{\mu\nu}$. The correspondence between the two is given in Eqs. (2.13), (3.12) and (3.13). However, one must be careful in doing the transformation since $\epsilon_3$ contains contributions to all $h_{11}$, $h_{22}$, and $h_{12}$. The exact correspondence is given by the following relation. After writing

$$\begin{aligned} &a\epsilon_1 + b\epsilon_2 + c\epsilon_3 + A\epsilon_1^2 + B\epsilon_2^2 + C\epsilon_3^2 \\ &\qquad + D\epsilon_1\epsilon_2 + E\epsilon_1\epsilon_3 + F\epsilon_2\epsilon_3, \end{aligned} \tag{3.37}$$

one can use the relationships between $\epsilon_i$ and $h_{\mu\nu}$ of Eq. (3.12) to obtain

$$\begin{aligned} &(\tfrac{a}{2} + \tfrac{c}{4})\, h_{11} + (\tfrac{b}{2} + \tfrac{c}{4})\, h_{22} + \tfrac{c}{4}\, h_{12} + \tfrac{c}{4}\, h_{21} \\ &+ (-\tfrac{a}{8} - \tfrac{c}{32} + \tfrac{A}{4} + \tfrac{C}{16} + \tfrac{E}{8})\, h_{11}^2 + (-\tfrac{b}{8} - \tfrac{c}{32} + \tfrac{B}{4} + \tfrac{C}{16} + \tfrac{F}{8})\, h_{22}^2 \\ &+ (-\tfrac{c}{16} + \tfrac{C}{8} + \tfrac{D}{4} + \tfrac{E}{8} + \tfrac{F}{8})\, h_{11}\, h_{22} + (-\tfrac{c}{16} + \tfrac{C}{8} + \tfrac{E}{8})\, h_{11}\, h_{12} \\ &+ (-\tfrac{c}{16} + \tfrac{C}{8} + \tfrac{F}{8})\, h_{11}\, h_{21} + (-\tfrac{c}{16} + \tfrac{C}{8} + \tfrac{F}{8})\, h_{22}\, h_{12} \\ &+ (-\tfrac{c}{16} + \tfrac{C}{8} + \tfrac{F}{8})\, h_{22}\, h_{21} + (-\tfrac{c}{32} + \tfrac{C}{16})\, h_{12}^2 + (-\tfrac{c}{16} + \tfrac{C}{8})\, h_{12}\, h_{21} \;. \end{aligned} \tag{3.38}$$

With the aid of the above equation, one can then easily rewrite the Feynman rules in terms of $h_{\mu\nu}\phi^2$. For the vertex $h_{11}(k)\phi(p)\phi(q)$ one has

$$\begin{aligned} &-4\,(1 - \tfrac{1}{2}\cos\tfrac{k_2}{2}) \sin(\tfrac{p_1}{2}) \sin(\tfrac{q_1}{2}) + 2 \cos(\tfrac{k_1}{2}) \sin(\tfrac{p_2}{2}) \sin(\tfrac{q_2}{2}) \\ &\qquad - 2\,(1 - \cos\tfrac{k_2}{2}) \sin(\tfrac{p_1+p_2}{2}) \sin(\tfrac{q_1+q_2}{2}) \;. \end{aligned} \tag{3.39}$$

When expanded out for small momenta it gives

$$\frac{1}{2}\,(-p_1\, q_1 + p_2\, q_2) = \frac{1}{2}\,(p\cdot q - 2\, p_1\, q_1) \;. \tag{3.40}$$

For the vertex $h_{22}(k)\phi(p)\phi(q)$ one obtains

$$\begin{aligned} &2 \cos(\tfrac{k_2}{2}) \sin(\tfrac{p_1}{2}) \sin(\tfrac{q_1}{2}) - 4\,(1 - \tfrac{1}{2}\cos\tfrac{k_1}{2}) \sin(\tfrac{p_2}{2}) \sin(\tfrac{q_2}{2}) \\ &\qquad - 2\,(1 - \cos\tfrac{k_1}{2}) \sin(\tfrac{p_1+p_2}{2}) \sin(\tfrac{q_1+q_2}{2}) \;, \end{aligned} \tag{3.41}$$

and its leading continuum approximation is given by

$$\frac{1}{2}\left(-p_2\, q_2 + p_1\, q_1\right) = \frac{1}{2}\left(p\cdot q - 2\, p_2\, q_2\right) \ . \tag{3.42}$$

For the vertex $h_{12}(k)\phi(p)\phi(q)$ ($= h_{21}(k)\phi(p)\phi(q)$ ) one has

$$\begin{aligned} &2\,\cos(\tfrac{k_2}{2})\,\sin(\tfrac{p_1}{2})\,\sin(\tfrac{q_1}{2}) + 2\,\cos(\tfrac{k_1}{2})\,\sin(\tfrac{p_2}{2})\,\sin(\tfrac{q_2}{2}) \\ &\qquad - 2\,\sin(\tfrac{p_1+p_2}{2})\,\sin(\tfrac{q_1+q_2}{2}) \ , \end{aligned} \tag{3.43}$$

and its leading continuum approximation is given by

$$-\frac{1}{2}\left(p_1\, q_2 + p_2\, q_1\right) \ . \tag{3.44}$$

Written in Lorentz covariant form, the leading continuum approximations for the vertices $h_{\mu\nu}(k)\phi(p)\phi(q)$ can be grouped together as

$$\frac{1}{2}\left(\delta_{\mu\nu} p\cdot q - p_\mu\, q_\nu - p_\nu\, q_\mu\right) \tag{3.45}$$

which is now identical to the usual continuum Feynman rule, derived from the original continuum action.

One can proceed in a similar way for the higher order vertices. The Feynman rules for the 2-scalar 2-graviton vertex written in the component forms are:
for the vertex $\epsilon_1(k)\epsilon_1(k')\phi(p)\phi(q)$,

$$\begin{aligned} &2\,\sin(\tfrac{p_1}{2})\,\sin(\tfrac{q_1}{2}) + \cos(\tfrac{k_1+k_2+k'_1+k'_2}{2})\,\sin(\tfrac{p_2}{2})\,\sin(\tfrac{q_2}{2}) \\ &\qquad - \tfrac{1}{2}\,\cos(\tfrac{k_2+k'_2}{2})\,\sin(\tfrac{p_1+p_2}{2})\,\sin(\tfrac{q_1+q_2}{2}) \ ; \end{aligned} \tag{3.46}$$

for the vertex $\epsilon_2(k)\epsilon_2(k')\phi(p)\phi(q)$,

$$\begin{aligned} &\cos(\tfrac{k_1+k_2+k'_1+k'_2}{2})\,\sin(\tfrac{p_1}{2})\,\sin(\tfrac{q_1}{2}) + 2\,\sin(\tfrac{p_2}{2})\,\sin(\tfrac{q_2}{2}) \\ &\qquad - \tfrac{1}{2}\cos(\tfrac{k_1+k'_1}{2})\,\sin(\tfrac{p_1+p_2}{2})\,\sin(\tfrac{q_1+q_2}{2}) \ ; \end{aligned} \tag{3.47}$$

for the vertex $\epsilon_3(k)\epsilon_3(k')\phi(p)\phi(q)$,

$$\begin{aligned} &3\,\cos(\tfrac{k_2+k'_2}{2})\,\sin(\tfrac{p_1}{2})\,\sin(\tfrac{q_1}{2}) + 3\,\cos(\tfrac{k_1+k'_1}{2})\,\sin(\tfrac{p_2}{2})\,\sin(\tfrac{q_2}{2}) \\ &\qquad - \sin(\tfrac{p_1+p_2}{2})\,\sin(\tfrac{q_1+q_2}{2}) \ ; \end{aligned} \tag{3.48}$$

for the vertex $\epsilon_1(k)\epsilon_2(k')\phi(p)\phi(q)$,

$$\begin{aligned} &2\,\cos(\tfrac{k'_1+k'_2}{2})\,\sin(\tfrac{p_1}{2})\,\sin(\tfrac{q_1}{2}) + 2\,\cos(\tfrac{k_1+k_2}{2})\,\sin(\tfrac{p_2}{2})\,\sin(\tfrac{q_2}{2}) \\ &\qquad - 2\,\cos(\tfrac{k_2-k'_1}{2})\,\sin(\tfrac{p_1+p_2}{2})\,\sin(\tfrac{q_1+q_2}{2}) \ ; \end{aligned} \tag{3.49}$$

for the vertex $\epsilon_1(k)\epsilon_3(k')\phi(p)\phi(q)$ ,

$$\begin{aligned} &-4\,\cos(\tfrac{k'_2}{2})\,\sin(\tfrac{p_1}{2})\,\sin(\tfrac{q_1}{2}) - 4\,\cos(\tfrac{k_1+k_2+k'_1}{2})\,\sin(\tfrac{p_2}{2})\,\sin(\tfrac{q_2}{2}) \\ &\qquad + 2\,\cos(\tfrac{k_2}{2})\,\sin(\tfrac{p_1+p_2}{2})\,\sin(\tfrac{q_1+q_2}{2}) \ ; \end{aligned} \tag{3.50}$$

for the vertex $\epsilon_2(k)\epsilon_3(k')\phi(p)\phi(q)$,

$$-4\,\cos(\tfrac{k_1+k_2+k_2'}{2})\,\sin(\tfrac{p_1}{2})\,\sin(\tfrac{q_1}{2}) - 4\,\cos(\tfrac{k_1'}{2})\,\sin(\tfrac{p_2}{2})\,\sin(\tfrac{q_2}{2}) + 2\,\cos(\tfrac{k_1}{2})\,\sin(\tfrac{p_1+p_2}{2})\,\sin(\tfrac{q_1+q_2}{2}). \tag{3.51}$$

After writing again the $\epsilon_i$ fields in terms of the $h_{\mu\nu}$ fields (see Eq. (3.12)), one obtains the Feynman rules for the vertex $h_{11}(k)h_{11}(k')\phi(p)\phi(q)$,

$$\begin{aligned}&(6-\tfrac{1}{2}\,\cos\tfrac{k_2}{2}+\tfrac{3}{2}\,\cos\tfrac{k_2+k_2'}{2}-4\,\cos\tfrac{k_2'}{2})\,\sin(\tfrac{p_1}{2})\,\sin(\tfrac{q_1}{2})\\ &+(\tfrac{3}{2}\,\cos\tfrac{k_1+k_1'}{2}-4\,\cos\tfrac{k_1'+k_1+k_2}{2}+2\,\cos\tfrac{k_1+k_2+k_1'+k_2'}{2}-\tfrac{1}{2}\cos\tfrac{k_1}{2})\,\sin(\tfrac{p_2}{2})\,\sin(\tfrac{q_2}{2})\\ &+(\cos\tfrac{k_2}{2}-\cos\tfrac{k_2+k_2'}{2})\,\sin(\tfrac{p_1+p_2}{2})\,\sin(\tfrac{q_1+q_2}{2})\ .\end{aligned} \tag{3.52}$$

The above expression should be symmetrized in $k$ and $k'$ to reflect the fact that an edge is shared between two neighboring triangles. For small momenta its leading continuum approximation is given by

$$\frac{1}{4}\,(-p\cdot q+4\,p_1q_1)\ . \tag{3.53}$$

For the vertex $h_{22}(k)h_{22}(k')\phi(p)\phi(q)$,

$$\begin{aligned}&(\tfrac{3}{2}\,\cos\tfrac{k_2+k_2'}{2}-4\,\cos\tfrac{k_1+k_2+k_2'}{2}+2\,\cos\tfrac{k_1+k_2+k_1'+k_2'}{2}-\tfrac{1}{2}\cos\tfrac{k_2}{2})\,\sin(\tfrac{p_1}{2})\,\sin(\tfrac{q_1}{2})\\ &+(6-4\,\cos\tfrac{k_1'}{2}+\tfrac{3}{2}\,\cos\tfrac{k_1+k_1'}{2}-\tfrac{1}{2}\cos\tfrac{k_1}{2})\,\sin(\tfrac{p_2}{2})\,\sin(\tfrac{q_2}{2})\\ &+(\cos\tfrac{k_1}{2}-\cos\tfrac{k_1+k_1'}{2})\,\sin(\tfrac{p_1+p_2}{2})\,\sin(\tfrac{q_1+q_2}{2})\ ,\end{aligned} \tag{3.54}$$

and its leading continuum approximation for small momenta is given by

$$\frac{1}{4}\,(-p\cdot q+4\,p_2q_2)\ . \tag{3.55}$$

For the vertex $h_{11}(k)h_{22}(k')\phi(p)\phi(q)$,

$$\begin{aligned}&2(-\cos\tfrac{k_2'}{2}+\cos\tfrac{k_1'+k_2'}{2}+\tfrac{3}{4}\,\cos\tfrac{k_2+k_2'}{2}-\cos\tfrac{k_1+k_2+k_2'}{2}-\tfrac{1}{4}\,\cos\tfrac{k_2}{2})\,\sin(\tfrac{p_1}{2})\,\sin(\tfrac{q_1}{2})\\ &+2(-\cos\tfrac{k_1'}{2}+\tfrac{3}{4}\,\cos\tfrac{k_1+k_1'}{2}+\cos\tfrac{k_1+k_2}{2}-\cos\tfrac{k_1+k_2+k_1'}{2}-\tfrac{1}{4}\,\cos\tfrac{k_1}{2})\,\sin(\tfrac{p_2}{2})\,\sin(\tfrac{q_2}{2})\\ &+(\cos\tfrac{k_1}{2}-2\,\cos\tfrac{k_1'-k_2}{2}+\cos\tfrac{k_2}{2})\,\sin(\tfrac{p_1+p_2}{2})\,\sin(\tfrac{q_1+q_2}{2})\ ,\end{aligned} \tag{3.56}$$

and its leading continuum approximation for small momenta is

$$-\frac{1}{4}\,p\cdot q\ . \tag{3.57}$$

For the vertex $h_{11}(k)h_{12}(k')\phi(p)\phi(q)$ ( $=h_{11}(k)h_{21}(k')\phi(p)\phi(q)$ ),

$$\begin{aligned}&\tfrac{1}{2}\,(-\cos\tfrac{k_2}{2}-4\,\cos\tfrac{k_2'}{2}+3\,\cos\tfrac{k_2+k_2'}{2})\,\sin(\tfrac{p_1}{2})\,\sin(\tfrac{q_1}{2})\\ &+\tfrac{1}{2}\,(-\cos\tfrac{k_1}{2}+3\,\cos\tfrac{k_1+k_1'}{2}-4\,\cos\tfrac{k_1+k_2+k_1'}{2})\,\sin(\tfrac{p_2}{2})\,\sin(\tfrac{q_2}{2})\\ &+\cos(\tfrac{k_2}{2})\,\sin(\tfrac{p_1+p_2}{2})\,\sin(\tfrac{q_1+q_2}{2})\ ,\end{aligned} \tag{3.58}$$

and its leading continuum approximation for small momenta is

$$\frac{1}{4}(p_1 q_2 + p_2 q_1) \ . \tag{3.59}$$

For the vertex $h_{22}(k)h_{12}(k')\phi(p)\phi(q)$ ( $= h_{22}(k)h_{21}(k')\phi(p)\phi(q)$ ),

$$\begin{aligned} &\tfrac{1}{2}(-\cos\tfrac{k_2}{2} + 3\cos\tfrac{k_2+k_2'}{2} - 4\cos\tfrac{k_1+k_2+k_2'}{2})\sin(\tfrac{p_1}{2})\sin(\tfrac{q_1}{2}) \\ &+ \tfrac{1}{2}(-\cos\tfrac{k_1}{2} - 4\cos\tfrac{k_1'}{2} + 3\cos\tfrac{k_1+k_1'}{2})\sin(\tfrac{p_2}{2})\sin(\tfrac{q_2}{2}) \\ &+ \cos(\tfrac{k_1}{2})\sin(\tfrac{p_1+p_2}{2})\sin(\tfrac{q_1+q_2}{2}) \ , \end{aligned} \tag{3.60}$$

and its leading continuum approximation for small momenta is

$$\frac{1}{4}(p_1 q_2 + p_2 q_1) \ . \tag{3.61}$$

For the vertex $h_{12}(k)h_{12}(k')\phi(p)\phi(q)$ ( $= h_{21}(k)h_{21}(k')\phi(p)\phi(q) = \frac{1}{2}[h_{12}(k)h_{21}(k')$ $\phi(p)\phi(q) + h_{21}(k)h_{12}(k')\phi(p)\phi(q)]$ ),

$$\begin{aligned} &\tfrac{1}{2}(-\cos\tfrac{k_2}{2} + 3\cos\tfrac{k_2+k_2'}{2})\sin(\tfrac{p_1}{2})\sin(\tfrac{q_1}{2}) \\ &+ \tfrac{1}{2}(-\cos\tfrac{k_1}{2} + 3\cos\tfrac{k_1+k_1'}{2})\sin(\tfrac{p_2}{2})\sin(\tfrac{q_2}{2}) \ , \end{aligned} \tag{3.62}$$

and its leading continuum approximation for small momenta is

$$\frac{1}{4} p \cdot q \ . \tag{3.63}$$

The above lattice Feynman rules for the 2-graviton 2-scalar vertex in terms of $h_{\mu\nu}$ can be compared to the continuum Feynman rules, which are given by the following expression

$$\begin{aligned} V_{\mu\nu,\alpha\beta} \ = \ & (1/4)\,[\,\delta_{\mu\alpha}(p_\nu q_\beta + p_\beta q_\nu) + \delta_{\mu\beta}(p_\nu q_\alpha + p_\alpha q_\nu) \\ & + \delta_{\nu\alpha}(p_\mu q_\beta + p_\beta q_\mu) + \delta_{\nu\beta}(p_\mu q_\alpha + p_\alpha q_\mu) - \delta_{\mu\nu}(p_\alpha q_\beta + p_\beta q_\alpha) \\ & - \delta_{\alpha\beta}(p_\mu q_\nu + p_\nu q_\mu) + (\delta_{\mu\nu}\delta_{\alpha\beta} - \delta_{\mu\alpha}\delta_{\nu\beta} - \delta_{\mu\beta}\delta_{\nu\alpha})\, p \cdot q\,] \ . \end{aligned} \tag{3.64}$$

Again one sees that the 2-graviton 2-scalar vertex in the leading continuum approximation reduces completely to the continuum Feynman rules.

In order to see how much the lattice Feynman rules depend on a particular lattice triangulation, one can derive the Feynman rules for the lattice obtained by a reflection with respect to the vertical axis (see Figure 6b). It turns out that these Feynman rules can simply be obtained by performing the following substitutions in all the above formulae for the vertices,

$$p_1 \to -p_1, q_1 \to -q_1, k_1 \to -k_1, k_1' \to -k_1' \ , \tag{3.65}$$

which indeed corresponds to a reflection about the vertical axis. In this case the leading continuum approximation of the lattice Feynman rules are again identical to the continuum Feynman rules. One therefore concludes that the two inequivalent lattice triangulations give the same physical results, at least to the first leading continuum order (for momenta which are small compared to the ultraviolet cutoff). This should not come as a surprise, since the two lattices correspond to two equivalent parameterizations of flat space, with an action that is parameterization invariant, at least for small deformations.

## 3.2 Conformal Anomaly

As an application, we will compute the graviton self-energy using the lattice Feynman rules developed above. There are two diagrams contributing to the lattice graviton self-energy, namely the vacuum polarization loop and the tadpole diagram shown in Figure 8.

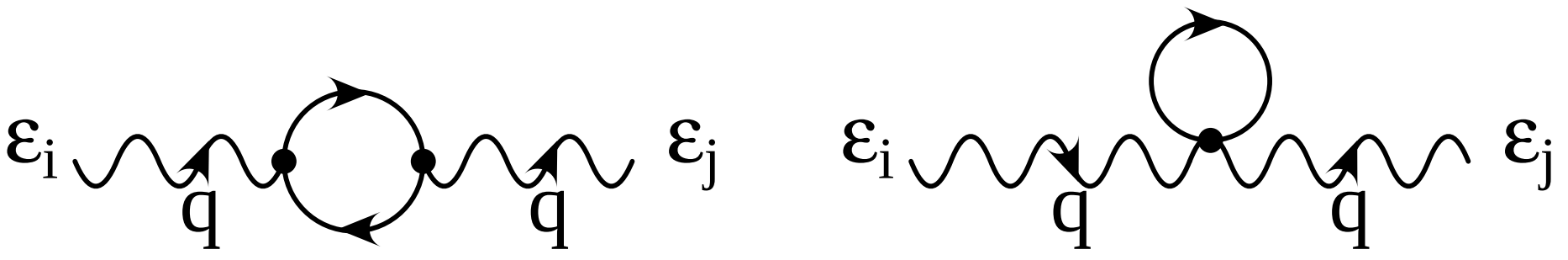

*Fig. 8. Lowest order diagrams contributing to the conformal anomaly.*

The evaluation of the above diagrams corresponds to a functional integration over the scalar field, performed to lowest order in the weak field expansion, and with a scalar field measure deriving from the functional measure

$$\int_0^\infty \prod_{\text{edges ij}} dl_{ij}^2 \; F_\epsilon[l] \times \int_{-\infty}^{\infty} \prod_{\text{sites i}} \prod_a d\phi_i^a \quad ; \tag{3.66}$$

the specific form of the gravitational measure will not matter in the following calculation since only the scalar field is integrated over. The relevant expressions for the vacuum polarization loop diagram, written in component form, are then

$$\begin{aligned} V_{\epsilon_1,\epsilon_1}(q) \;=\; & \frac{1}{2}\int_{-\pi}^{\pi}\int_{-\pi}^{\pi}\frac{d^2p}{(2\,\pi)^2}\Big\{-2\,\sin\frac{p_1}{2}\,\sin\frac{p_1+q_1}{2} \\ & +\cos\frac{q_2}{2}\,\sin\frac{p_1+p_2}{2}\,\sin\frac{p_1+q_1+p_2+q_2}{2}\Big\}^2 \\ & \times\frac{1}{16\,[\,\sum_\mu\,\sin^2(\frac{p_\mu}{2})\,][\,\sum_\mu\,\sin^2(\frac{p_\mu+q_\mu}{2})\,]}, \end{aligned} \tag{3.67}$$

$$\begin{aligned} V_{\epsilon_2,\epsilon_2}(q) \;=\; & \frac{1}{2}\int_{-\pi}^{\pi}\int_{-\pi}^{\pi}\frac{d^2p}{(2\,\pi)^2}\Big\{-2\,\sin\frac{p_2}{2}\,\sin\frac{p_2+q_2}{2} \\ & +\cos\frac{q_1}{2}\,\sin\frac{p_1+p_2}{2}\,\sin\frac{p_1+q_1+p_2+q_2}{2}\Big\}^2 \\ & \times\frac{1}{16\,[\,\sum_\mu\,\sin^2(\frac{p_\mu}{2})\,][\,\sum_\mu\,\sin^2(\frac{p_\mu+q_\mu}{2})\,]}, \end{aligned} \tag{3.68}$$

$$
\begin{aligned}
V_{\epsilon_3,\epsilon_3}(q) &= \frac{1}{2}\int_{-\pi}^{\pi}\int_{-\pi}^{\pi}\frac{d^2p}{(2\,\pi)^2}\left\{2\,\cos\frac{q_2}{2}\,\sin\frac{p_1}{2}\,\sin\frac{p_1+q_1}{2}\right. \\
&\left. +2\,\cos\frac{q_1}{2}\,\sin\frac{p_2}{2}\sin\frac{p_2+q_2}{2} - 2\,\sin\frac{p_1+p_2}{2}\sin\frac{p_1+q_1+p_2+q_2}{2}\right\}^2 \\
&\times\frac{1}{16\,[\,\sum_\mu\,\sin^2(\frac{p_\mu}{2})\,][\,\sum_\mu\,\sin^2(\frac{p_\mu+q_\mu}{2})\,]},
\end{aligned} \tag{3.69}
$$

$$
\begin{aligned}
V_{\epsilon_1,\epsilon_2}(q) &= \frac{1}{2}\int_{-\pi}^{\pi}\int_{-\pi}^{\pi}\frac{d^2p}{(2\,\pi)^2}\left\{(-2\,\sin\frac{p_1}{2}\,\sin\frac{p_1+q_1}{2} + \cos\frac{q_2}{2}\,\sin\frac{p_1+p_2}{2}\right. \\
&\times\sin\frac{p_1+q_1+p_2+q_2}{2})\times(-2\,\sin\frac{p_2}{2}\,\sin\frac{p_2+q_2}{2} + \\
&\left.\cos\frac{q_1}{2}\,\sin\frac{p_1+p_2}{2}\,\sin\frac{p_1+q_1+p_2+q_2}{2})\right\} \\
&\times\frac{1}{16\,[\,\sum_\mu\,\sin^2(\frac{p_\mu}{2})\,][\,\sum_\mu\,\sin^2(\frac{p_\mu+q_\mu}{2})\,]}.
\end{aligned} \tag{3.70}
$$

$$
\begin{aligned}
V_{\epsilon_1,\epsilon_3}(q) &= \frac{1}{2}\int_{-\pi}^{\pi}\int_{-\pi}^{\pi}\frac{d^2p}{(2\,\pi)^2}\left\{(-2\,\sin\frac{p_1}{2}\,\sin\frac{p_1+q_1}{2} + \cos\frac{q_2}{2}\,\sin\frac{p_1+p_2}{2}\right. \\
&\times\sin\frac{p_1+q_1+p_2+q_2}{2})\times(2\,\cos\frac{q_2}{2}\,\sin\frac{p_1}{2}\,\sin\frac{p_1+q_1}{2} + \\
&\left.2\,\cos\frac{q_1}{2}\,\sin\frac{p_2}{2}\,\sin\frac{p_2+q_2}{2} - 2\,\sin\frac{p_1+p_2}{2}\,\sin\frac{p_1+q_1+p_2+q_2}{2}\,)\right\} \\
&\times\frac{1}{16\,[\,\sum_\mu\,\sin^2(\frac{p_\mu}{2})\,][\,\sum_\mu\,\sin^2(\frac{p_\mu+q_\mu}{2})\,]}.
\end{aligned} \tag{3.71}
$$

$$
\begin{aligned}
V_{\epsilon_2,\epsilon_3}(q) &= \frac{1}{2}\int_{-\pi}^{\pi}\int_{-\pi}^{\pi}\frac{d^2p}{(2\,\pi)^2}\left\{(-2\,\sin\frac{p_2}{2}\,\sin\frac{p_2+q_2}{2} + \cos\frac{q_1}{2}\,\sin\frac{p_1+p_2}{2}\right. \\
&\times\sin\frac{p_1+q_1+p_2+q_2}{2})\times(2\,\cos\frac{q_2}{2}\,\sin\frac{p_1}{2}\,\sin\frac{p_1+q_1}{2} + \\
&\left.2\,\cos\frac{q_1}{2}\,\sin\frac{p_2}{2}\,\sin\frac{p_2+q_2}{2} - 2\,\sin\frac{p_1+p_2}{2}\,\sin\frac{p_1+q_1+p_2+q_2}{2}\,)\right\} \\
&\times\frac{1}{16\,[\,\sum_\mu\,\sin^2(\frac{p_\mu}{2})\,][\,\sum_\mu\,\sin^2(\frac{p_\mu+q_\mu}{2})\,]}.
\end{aligned} \tag{3.72}
$$

For zero external momentum $q$ all the quantities $V_{\epsilon_i,\epsilon_j}$'s are infrared finite, and one finds

$$
\begin{aligned}
V_{\epsilon_1,\epsilon_1}(q=0) &= \frac{1}{16}\left(1-\frac{2}{\pi}\right) \\
V_{\epsilon_2,\epsilon_2}(q=0) &= \frac{1}{16}\left(1-\frac{2}{\pi}\right) \\
V_{\epsilon_3,\epsilon_3}(q=0) &= \frac{1}{8}\left(1-\frac{2}{\pi}\right)
\end{aligned}
$$

$$
\begin{aligned}
V_{\epsilon_1,\epsilon_2}(q=0) &= 0 \\
V_{\epsilon_1,\epsilon_3}(q=0) &= -\frac{1}{16}\left(-1+\frac{1}{\pi}\right) \\
V_{\epsilon_2,\epsilon_3}(q=0) &= -\frac{1}{16}\left(-1+\frac{1}{\pi}\right)
\end{aligned}
\tag{3.73}
$$

The $q^2$-dependent terms are the ones that are of physical importance. Expanding the lattice integrals for small external momentum one obtains for example

$$
V_{\epsilon_1,\epsilon_1}(q1=q2=q) = \frac{1}{16}\left(1-\frac{2}{\pi}\right) + \frac{1}{96\pi}q^2 + O(q^4) \ , \tag{3.74}
$$

which can be compared with the continuum result for the metric vacuum polarization

$$
\Pi_{1111} = \frac{1}{96\pi}q^2 + O(q^4) \ . \tag{3.75}
$$

A complete analytic evaluation of all the lattice integrals in the limit of small momentum is beyond the scope of this paper, and will be presented elsewhere.

The expressions relevant for the tadpole diagram, written in component form, are

$$
\begin{aligned}
T_{\epsilon_1,\epsilon_1}(q) &= \int_{-\pi}^{\pi}\int_{-\pi}^{\pi}\frac{d^2p}{(2\,\pi)^2}\left\{2\,\sin^2(\frac{p_1}{2}) + \sin^2(\frac{p_2}{2}) - \frac{1}{2}\,\sin^2(\frac{p_1+p_2}{2})\right\} \\
&\quad \times \frac{1}{4\,[\sum_\mu \sin^2(\frac{p_\mu}{2})]}.
\end{aligned}
\tag{3.76}
$$

$$
\begin{aligned}
T_{\epsilon_2,\epsilon_2}(q) &= \int_{-\pi}^{\pi}\int_{-\pi}^{\pi}\frac{d^2p}{(2\,\pi)^2}\left\{\sin^2(\frac{p_1}{2}) + 2\,\sin^2(\frac{p_2}{2}) - \frac{1}{2}\,\sin^2(\frac{p_1+p_2}{2})\right\} \\
&\quad \times \frac{1}{4\,[\sum_\mu \sin^2(\frac{p_\mu}{2})]}.
\end{aligned}
\tag{3.77}
$$

$$
\begin{aligned}
T_{\epsilon_3,\epsilon_3}(q) &= \int_{-\pi}^{\pi}\int_{-\pi}^{\pi}\frac{d^2p}{(2\,\pi)^2}\left\{3\,\sin^2(\frac{p_1}{2}) + 3\,\sin^2(\frac{p_2}{2}) - \sin^2(\frac{p_1+p_2}{2})\right\} \\
&\quad \times \frac{1}{4\,[\sum_\mu \sin^2(\frac{p_\mu}{2})]}.
\end{aligned}
\tag{3.78}
$$

$$
\begin{aligned}
T_{\epsilon_1,\epsilon_2}(q) &= \int_{-\pi}^{\pi}\int_{-\pi}^{\pi}\frac{d^2p}{(2\,\pi)^2}\,\cos\frac{q_1+q_2}{2}\,\left\{2\,\sin^2(\frac{p_1}{2}) + 2\,\sin^2(\frac{p_2}{2})\right. \\
&\quad \left. -2\,\sin^2(\frac{p_1+p_2}{2})\right\} \times \frac{1}{4\,[\sum_\mu \sin^2(\frac{p_\mu}{2})]}.
\end{aligned}
\tag{3.79}
$$

$$
\begin{aligned}
T_{\epsilon_1,\epsilon_3}(q) &= \int_{-\pi}^{\pi}\int_{-\pi}^{\pi}\frac{d^2p}{(2\,\pi)^2}\,\cos\frac{q_2}{2}\left\{-4\,\sin^2(\frac{p_1}{2}) - 4\,\sin^2(\frac{p_2}{2}) + 2\,\sin^2(\frac{p_1+p_2}{2})\right\} \\
&\quad \times \frac{1}{4\,[\sum_\mu \sin^2(\frac{p_\mu}{2})]}.
\end{aligned}
\tag{3.80}
$$

$$
\begin{aligned}
T_{\epsilon_2,\epsilon_3}(q) \;&=\; \int_{-\pi}^{\pi}\int_{-\pi}^{\pi} \frac{d^2p}{(2\,\pi)^2}\,\cos\frac{q_1}{2}\left\{-4\,\sin^2(\frac{p_1}{2}) - 4\,\sin^2(\frac{p_2}{2}) + 2\,\sin^2(\frac{p_1+p_2}{2})\right\} \\
&\times \frac{1}{4\,[\sum_\mu \sin^2(\frac{p_\mu}{2})]} . 
\end{aligned}
\tag{3.81}
$$

The first three integrals are $q$-independent, while in the last three the $q$-dependence factorizes. One finds

$$
\begin{aligned}
T_{\epsilon_1,\epsilon_1}(q) &= \frac{3}{8} - \frac{1}{4\pi} \\
T_{\epsilon_2,\epsilon_2}(q) &= \frac{3}{8} - \frac{1}{4\pi} \\
T_{\epsilon_3,\epsilon_3}(q) &= \frac{3}{4} - \frac{1}{2\pi} \\
T_{\epsilon_1,\epsilon_2}(q) &= \cos(\frac{q_1+q_2}{2})\left(\frac{1}{2} - \frac{1}{\pi}\right) \\
T_{\epsilon_1,\epsilon_3}(q) &= \cos(\frac{q_2}{2})\left(-1 + \frac{1}{\pi}\right) \\
T_{\epsilon_2,\epsilon_3}(q) &= \cos(\frac{q_1}{2})\left(-1 + \frac{1}{\pi}\right)
\end{aligned}
\tag{3.82}
$$

The terms in $T_{\epsilon_i,\epsilon_j}(q)$ are required to cancel some of the unwanted, including noncovariant, terms in $V_{\epsilon_i,\epsilon_j}(q)$. Also, a number of contributions to the lattice vacuum polarization can be shown to contribute to a renormalization of the cosmological constant. It should be noted here that the lattice form of the cosmological constant term (which just corresponds to the total area of the triangulated manifold) contains, in contrast to the continuum, momentum dependent terms [9]. The reason for this is that the lattice area terms couple neighboring edges, and lead therefore to some residual local interactions between the edges variables, as shown in Eq. (3.14).

Eventually one needs to rotate the final answer for the vacuum polarization from the $\epsilon_i$ to the $h_{\mu\nu}$ variables, which is achieved to linear order (and for small $q$) by the matrix

$$
V = \begin{pmatrix} \frac{1}{2} & 0 & 0 \\ 0 & \frac{1}{2} & 0 \\ \frac{1}{4} & \frac{1}{4} & \frac{1}{2} \end{pmatrix} \quad . \tag{3.83}
$$

with $\epsilon(q) = V \cdot h(q)$ (see Eq. (3.13)). Since the integration over the scalar gives

$$
I_{eff} \;=\; -\tfrac{1}{2}\int \frac{d^2q}{(2\pi)^2}\, h(q)\,\cdot \Pi(h)\,\cdot h(q) \;=\; -\tfrac{1}{2}\int \frac{d^2q}{(2\pi)^2}\,\epsilon(q)\cdot\Pi(\epsilon)\cdot\epsilon(q) \;\; , \tag{3.84}
$$

one obtains the correspondence between the two polarization quantities,

$$
\Pi(h) \;=\; V^{-1}\cdot\Pi(\epsilon)\cdot V \;\; , \tag{3.85}
$$

correct to lowest order in the weak field expansion.

Analytical evaluations of the remaining integrals in terms of Bessel functions will be presented elsewhere [34]. In the following we will calculate the graviton self-energy using the lattice Feynman rules in the leading continuum order, by doing a continuum approximation to the integrands valid for small momenta. The procedure is justified for integrands that are sharply peaked in the low momentum region, and neglects the effects due to presence of a high momentum cutoff.

The calculation is most easily done by using the Feynman rules in terms of $h_{\mu\nu}$ in the Lorentz covariant form, which were given before. The vacuum polarization loop contributing to the graviton self-energy then reduces to

$$\begin{aligned} V_{\mu\nu,\alpha\beta}(q) &= \frac{1}{2} \int \frac{d^2 p}{(2\,\pi)^2} \frac{t_{\mu\nu}(p,q)\, t_{\alpha\beta}(p,q)}{p^2\,(p+q)^2} \\ t_{\mu\nu}(p,q) &= \frac{1}{2} \left[ \delta_{\mu\nu} p \cdot (p+q) - p_\mu\,(p_\nu + q_\nu) - p_\nu\,(p_\mu + q_\mu) \right] \; . \end{aligned} \tag{3.86}$$

The calculation of the integral is easily done using dimensional regularization. One obtains

$$V_{\mu\nu,\alpha\beta}(q) = \frac{1}{48\,\pi} \left( q^2 \delta_{\mu\nu} - q_\mu q_\nu \right) \frac{1}{q^2} \left( q^2 \delta_{\alpha\beta} - q_\alpha q_\beta \right) \; . \tag{3.87}$$

As expected, the tadpole contribution $T_{\mu\nu,\alpha\beta}(q)$ to the graviton self-energy is 0. The remaining graviton self-energy contribution is given by

$$\begin{aligned} \Pi_{\mu\nu,\alpha\beta}(q) &= V_{\mu\nu,\alpha\beta}(q) + T_{\mu\nu,\alpha\beta}(q) \\ &= \frac{1}{48\,\pi} \left( q^2 \delta_{\mu\nu} - q_\mu q_\nu \right) \frac{1}{q^2} \left( q^2 \delta_{\alpha\beta} - q_\alpha q_\beta \right) \; . \end{aligned} \tag{3.88}$$

The above result is valid for one real massless scalar field. For a $D$-component scalar field, the above result would get multiplied by a factor of $D$. Then the effective action, obtained from integrating out the scalar degrees of freedom, and to lowest order in the weak field expansion, is given by

$$I_{eff} \; = \; -\tfrac{1}{2} \int \frac{d^2 q}{(2\pi)^2} \, h_{\mu\nu}(q) \, \Pi_{\mu\nu\rho\sigma}(q) \, h_{\rho\sigma}(-q) \; , \tag{3.89}$$

with the scalar vacuum polarization given by

$$\Pi_{\mu\nu\rho\sigma}(q) = \frac{D}{48\pi} \left( q_\mu q_\nu - \delta_{\mu\nu} q^2 \right) \frac{1}{q^2} \left( q_\rho q_\sigma - \delta_{\rho\sigma} q^2 \right) \; . \tag{3.90}$$

In the lattice analog of the continuum conformal gauge,

$$g_{\mu\nu}(x) \; = \; \delta_{\mu\nu} \, e^{\varphi(x)} \; , \tag{3.91}$$

one can write for the scalar curvature

$$(q_\mu q_\nu - \delta_{\mu\nu} q^2) h_{\mu\nu}(q) = R(q) = q^2 \varphi(q) \; , \tag{3.92}$$

and therefore re-write the effective action in the form

$$I_{eff}(\varphi) \; = \; -\frac{D}{96\pi} \int \frac{d^2 q}{(2\pi)^2} \, \varphi(q) \, q^2 \varphi(-q) \; , \tag{3.93}$$

which in real space becomes the well-known Liouville action

$$I_{eff}(\varphi) \; = \; -\frac{D}{96\pi} \int d^2x \left[ (\partial_\mu \varphi)^2 + (\lambda - \lambda_c) \, e^{\varphi} \right] \quad . \tag{3.94}$$

One therefore completely recovers the result derived perturbatively from the continuum Feynman rules, as given for example in [15]).

In the continuum, the effective action term of Eq. (3.94) arising from the conformal anomaly can be written in covariant form as

$$\tfrac{1}{2} \int d^2x \, d^2y \, R \sqrt{g}(x) \, \langle x | \frac{1}{-\partial^2 + m^2} | y \rangle \, R \sqrt{g}(y) \quad , \tag{3.95}$$

where $\partial^2$ is the continuum covariant Laplacian, $\partial^2 \equiv \partial_\mu (\sqrt{g} g^{\mu\nu} \partial_\nu)$, and $m^2 \rightarrow 0$. In two dimensions and for flat space, $\langle x | \frac{1}{\partial^2} | y \rangle \sim \frac{1}{2\pi} \log |x - y|$. Using the correspondence between lattice and continuum curvature operators derived in [5], its lattice form gives rise to an effective long-range interaction between deficit angles of the type

$$\tfrac{1}{2} \sum_{\mathrm{hinges} h, h'} \delta_h \, \left[ \frac{1}{-\Delta + m^2} \right]_{h,h'} \delta_{h'} \quad , \tag{3.96}$$

where $\Delta$ is the nearest-neighbor covariant lattice Laplacian, as obtained from the discrete scalar action, and $m^2 \leftarrow 0$ an infrared mass regulator. This result was given in [9]; see also the recent discussion in [35].

In general one cannot expect the lattice expression for the vacuum polarization to match completely the continuum expression already at the lowest order of perturbation theory. Since there is *no* small parameter controlling the weak field expansion in two dimensions, it is difficult to see why the first few orders on the lattice should suffice. Indeed in the weak field expansion the background lattice is usually taken to be regular, which leads to a set of somewhat preferred lattice directions for high momenta, which are close to the momentum cutoff at $\pm\pi$. It would seem though that this is an artifact of the choice of background lattice (which is necessarily rigid), and whose effects are eventually washed out when the fluctuations is the edge lengths are properly accounted for in higher order in the weak field expansion. The fact that the conformal mode in fact remains massless in two dimensions in the full numerical, non-perturbative treatment of the lattice theory was shown in [16,28], without the necessity of any sort of fine-tuning of bare parameters.

The gravitational contribution to the effective action in the lattice conformal gauge can, at least in principle, be computed in a similar way. Let us sketch here how the analogous lattice calculation would proceed; a more detailed discussion of the relevant calculations will be presented elsewhere. In the continuum the metric perturbations are naturally decomposed into conformal and diffeomorphism parts,

$$\delta g_{\mu\nu}(x) = g_{\mu\nu}(x) \, \delta\varphi(x) + \nabla_\mu \chi_\nu(x) + \nabla_\nu \chi_\mu(x) \quad . \tag{3.97}$$

where $\nabla_\nu$ denotes the covariant derivative. A similar decomposition can be done for the lattice degrees of freedom, by separating out the lattice gauge transformations [2] (which act on the vertices and change the edge lengths without changing the local curvatures) from the conformal transformations (which change the local volumes and curvatures) in Eq. (3.20). The explicit form for the lattice diffeomorphisms, to lowest order in the lattice weak field expansion, is given in Eq. (3.18), which makes it obvious that such a decomposition can indeed be performed on the lattice. After rewriting the gravitational functional measure in terms of conformal and diffeomorphism degrees of freedom,

$$[dg] \;=\; [d\varphi][d\chi]\,[\det(L^{+}L)]^{\frac{1}{2}} \quad , \tag{3.98}$$

which involves the Jacobian of the operator $L$, which in the continuum is determined from

$$(L^{+}L\;\chi)_\mu = \nabla^\nu(\nabla_\mu\chi_\nu + \nabla_\nu\chi_\mu - g_{\mu\nu}\nabla^\rho\chi_\rho) \;\; , \tag{3.99}$$

one has for the effective action contribution in the conformal gauge, and to lowest order in the weak field expansion,

$$[\det(L^{+}L)]^{-\frac{1}{2}} \;\sim\; \exp\left\{-I_{eff}(\varphi)\right\} \;\; . \tag{3.100}$$

The lattice form of $L$ depends on the specific form of the lattice gauge fixing term. On the lattice the functional integration is performed over the edge lengths, see Eq. (2.3). The lattice conformal gauge choice corresponds to an assignment of edge lengths such that

$$g_{ij}(n) = \begin{pmatrix} l_3^2 & \frac{1}{2}(-l_1^2+l_2^2+l_3^2) \\ \frac{1}{2}(-l_1^2+l_2^2+l_3^2) & l_2^2 \end{pmatrix} \;\approx\; \delta_{ij}\; e^{\varphi(n)} \;\; . \tag{3.101}$$

The lattice fields $\varphi(n)$ are defined on the lattice vertices. So are the gauge degrees of freedom $\chi_\mu(n)$, as can be seen from Eq. (3.19). It should therefore be clear that the choice of lattice conformal gauge corresponds to a re-assignment of edge lengths about each vertex which leaves the local curvature unchanged but brings the induced metric into diagonal form; it corresponds to a choice of approximately right-angle triangles at each vertex [32].

A diagrammatic calculation, similar to the one for the scalar field contribution, gives in the continuum the celebrated result [15]

$$\Pi_{\mu\nu\rho\sigma}(q) \;=\; \frac{13}{48\pi}\,(q_\mu q_\nu - \delta_{\mu\nu}q^2)\,\frac{1}{q^2}\,(q_\rho q_\sigma - \delta_{\rho\sigma}q^2) \;\; . \tag{3.102}$$

As a consequence the total Liouville action for the field $\varphi = \frac{1}{\partial^2}R$ becomes

$$I_{eff}(\varphi) \;=\; \frac{26-D}{96\pi}\int d^2x\,\left[(\partial_\mu\varphi)^2 + (\lambda-\lambda_c)\,e^{\varphi}\right] \quad , \tag{3.103}$$

Thus to lowest order in the weak field expansion the critical value of $D$ for which the action vanishes is $D_c = 26$, but this number is expected to get modified by higher order quantum corrections. In any case, for sufficiently large $D$ one expects an instability to develop. Numerical nonperturbative studies of two-dimensional gravity suggest that in the lattice theory the correction is large, and one finds that the threshold of instability moves to $D_c \approx 13$ [28]. It is unclear if this critical value can be regarded as truly universal, and independent for example on the detailed choice of gravitational measure.

# 4 Conclusions

In the previous section we have presented results relevant for the weak field expansion of lattice gravity. We have shown the precise correspondence between continuum and lattice degrees of freedom within the context of such an expansion. To avoid unnecessary technical complications, most of our discussion has focused on the two-dimensional case, but the methods presented in this paper can be applied to any dimension. In the purely gravitational case, we have shown that the presence of zero modes is tied to the existence of a local invariance of the gravitational action. This invariance corresponds precisely to the diffeomorphisms in the continuum, with appropriate deformations in the edge lengths playing the role of local gauge variations of the metric in the continuum.

We have then derived the Feynman rules for gravity coupled to a massless scalar field, to lowest order in the weak field expansion. Although the lattice Feynman rules for the edge length vertices appear to be rather unwieldy, they actually reduce to the familiar continuum form when re-expressed in terms of the weak field metric field, and we have presented this important result in detailed form. As an application, the two-dimensional conformal anomaly due to a massless scalar field was computed by diagrammatic methods. We have given explicitly the relevant lattice integrals, which involve among other things a new diagram, the tadpole term, which vanishes in the continuum but is necessary on the lattice for canceling unwanted terms. Finally we have shown that in the leading continuum approximation the expected continuum form for the anomaly is obtained, with the correct coefficient.

The procedure followed here in deriving the Feynman rules for lattice gravity works in any dimension, including therefore the physical case of four dimensions. The lack of perturbative renormalizability in four dimensions is not ameliorated though by the presence of an explicit lattice cutoff, and non-perturbative searches for an ultraviolet fixed point and a lattice continuum limit are still required.

**Acknowledgements**

One of the authors (HWH) would like to thank G. Veneziano and the Theory Division at CERN for hospitality during the initial stages of this work.